\documentclass[pra,aps,twocolumn,showpacs,amsmath,,amsfonts,amssymb,superscriptaddress]{revtex4}
\usepackage{graphicx,color}

\definecolor{darkblue}{rgb}{0,0,0.75}
\definecolor{darkgreen}{rgb}{0.,0.7,0}
\definecolor{darkred}{rgb}{0.7,0,0}
\definecolor{note}{rgb}{0.05,0,0.75}
\definecolor{note2}{rgb}{0.2,0.5,0.0}
\usepackage[unicode, colorlinks, citecolor=darkblue, linkcolor=darkred, urlcolor=blue]{hyperref}

\newcommand{\popl}[1]{{\color{note2} #1}}

\begin{document}

\title{On fundamental diffraction limitation of finesse of a Fabry-Perot cavity}

\author{Mikhail V. Poplavskiy}
\affiliation{Faculty of Physics, M.V.Lomonosov Moscow State University, Moscow 119991, Russia}

\author{Andrey B. Matsko}
\affiliation{OEwaves Inc., 465 North Halstead Street, Suite 140, Pasadena, CA 91107, USA}

\author{Hiroaki Yamamoto}
\affiliation{LIGO Laboratory, California Institute of Technology, MC~100-36, Pasadena, CA 91125, USA}

\author{Sergey P. Vyatchanin}
\affiliation{Faculty of Physics, M.V.Lomonosov Moscow State University, Moscow 119991, Russia}

\begin{abstract}
We perform a theoretical study of finesse limitations of a Fabry-Perot (FP) cavity occurring due to finite size, asymmetry, as well as imperfections  of the cavity mirrors. A method of numerical simulations of the eigenvalue problem applicable for both the fundamental and high order cavity modes is suggested. Using this technique we find spatial profile of the modes and their round-trip diffraction loss. The results of the numerical simulations and analytical calculations are nearly identical when we consider a conventional FP cavity.  The proposed numerical technique has much broader applicability range and is valid for any FP cavity with arbitrary non-spherical mirrors which have cylindrical symmetry but disturbed in an asymmetric way, for example, by tilt or roughness of their mirrors.
\end{abstract}

\pacs{95.55.Ym, 42.60.Da, 42.79.Bh, 42.65.Sf}

\maketitle

\section{Introduction}

A Fabry-Perot (FP) cavity is one of the best known physical objects in linear optics covered in multiple textbooks \cite{Siegman1986, Saleh1991, Marcuse1972, Svelto2010}.  At the simplest level it is considered as a single dimension (1D) structure having two mirrors characterized with integral power transmission $T$ and attenuation $R$. A solution of a 1D wave equation with the boundary conditions taking into account the mirror properties describes the structure completely. This consideration, though, is not very accurate is applied to a realistic system. A stricter analysis of the cavity is more involved. It requires consideration of the finite size and shape of the cavity mirrors and calls for a two dimensional (2D) model that takes diffraction into account \cite{Siegman1986, Marcuse1972}. The attenuation frequently slips away from the 2D analysis and is introduced by hand in a way similar to the oversimplified 1D picture. The question about optimization of the cavity to reduce the attenuation and achieve 
the highest possible finesse was not studied in detail, especially for the realistic cases of slightly non-symmetric cavities. The sources of loss were identified  and evaluated numerically \cite{92pare,05kuznetsov,07tiffani}, but no in-depth investigation was performed. In this paper we fill the gap and report on the detailed analytical and numerical investigation of the FP cavity loss factors based on the fundamental wave optics effects.

Two spherical mirrors having the same symmetry axis and separated by a macroscopic distance is the simplest and well known optical model of a FP cavity. Gaussian beam is formally an accurate presentation of the modes in a FP cavity with infinite in transversal direction spherical mirrors. For a cavity with finite sized spherical mirrors Gaussian beam still remains a good approximation of the spatial distribution of the power in the cavity modes, as confirmed by numerical simulations. This assumption allows estimating analytically the diffraction loss. It can be done by evaluating the relative part of the incident light power that is not reflected by the mirror, using so called clip approximation \cite{hercher1968,Siegman1986, fulda2010}. We can also apply the clip approximation in order to estimate analytically the loss appearing due to small tilt of the mirrors \cite{hauck1980,hefetz1997} or inhomogeneous thermal heating of the mirror surface \cite{vinet2009}. On the other hand, this approximation is not 
accurate for higher order optical modes (HOOM) and only numerical calculations help in this case.

Realistic FP cavities have a large density of high finesse modes and the fundamental mode family, characterized with a single intensity peak in the plain orthogonal to the cavity axis, is one of them \cite{BoydBSTJ1961,Siegman1986}. It is possible to reduce the observable spectral density of modes by optimizing the input-output optics preventing excitation of the high-order modes of a linear cavity. However, the intrinsic multimode spectrum is detrimental in multiple nonlinear applications. High-order modes lead to mode competition in lasers \cite{Sargent1974} as well as unwanted nonlinear, e.g. opto-mechanical, instabilities. The later ones are observed, for instance, in gravitational wave detectors such as Advanced LIGO (aLIGO) \cite{aLIGO2013,aLIGO2014, aLIGO2014b} where high intracavity optical power leads to the excitation of the mechanical modes of the mirrors and generation of associated optical harmonics localized in the high-order optical cavity modes \cite{01a1BrStVy,02a1BrStVy}. Reduction of the 
optical
spectral density suppresses the process \cite{14a1FeDeVyMa, 16a1MaPoYaVy}. An accurate analytical description of the modes of the cavities with non-spherical mirrors does not exist and numerical modeling is essential to find the eigenfrequencies, attenuation, and field profile of the cavity modes. The proposed here approach helps solving the problem.

The are several algorithms for numerical solution of the eigenvalue problem of a FP cavity with non-spherical mirrors. The numerical simulations using 2D or 3D models call for a significant computing time and do not converge fast enough  \cite{prazeres1992,hong2011,benedikter2015}. It was shown that the eigenvalue problem of a 3D FP cavity with arbitrary shaped mirrors with axial symmetry can be reduced to 1D model \cite{93vinet} by application of Hankel Transform for computation of {\em axial symmetric} modes and a Matlab code is freely available \cite{10yamamoto}. In this paper we generalize this method for non-axially symmetric HOOM  with dependence $e^{i\ell\phi}$ on azimuthal angle $\phi$ (integer $\ell$ is non-zero). We apply this method for calculation of normal modes of a FP cavity with non-spherical (but axially symmetric) mirrors  and its diffraction losses. Furthermore, we apply a successive approximation method to evaluate the diffraction loss produced by small shape perturbations and tilt of a 
cavity mirror utilizing results of 1D numerical calculations considering it as a zero-order approximation. While we utilize aLIGO cavity parameters in our simulation, our analysis is valid for any type of a FP cavity with loss limited by diffraction, mirror misalignment as well as imperfections.

The paper is organized as follows. We describe a physical model of a 3D FP cavity and formulate the associate eigenvalue problem in Section II. An analytical model of the FP with spherical mirrors is presented in Section III. To obtain numerical estimate for the loss of the modes we use a well tabulated example of LIGO cavity. Numerical simulation method of finding loss of a FP cavity having axial symmetry is described in Section IV. In Sections V and VI we describe an analytical approach of evaluating finesse of an asymmetric FP cavity with tilted mirror as well as rough mirror surface.

\section{Model}\label{model}

\begin{figure}
 \includegraphics[width=0.35\textwidth]{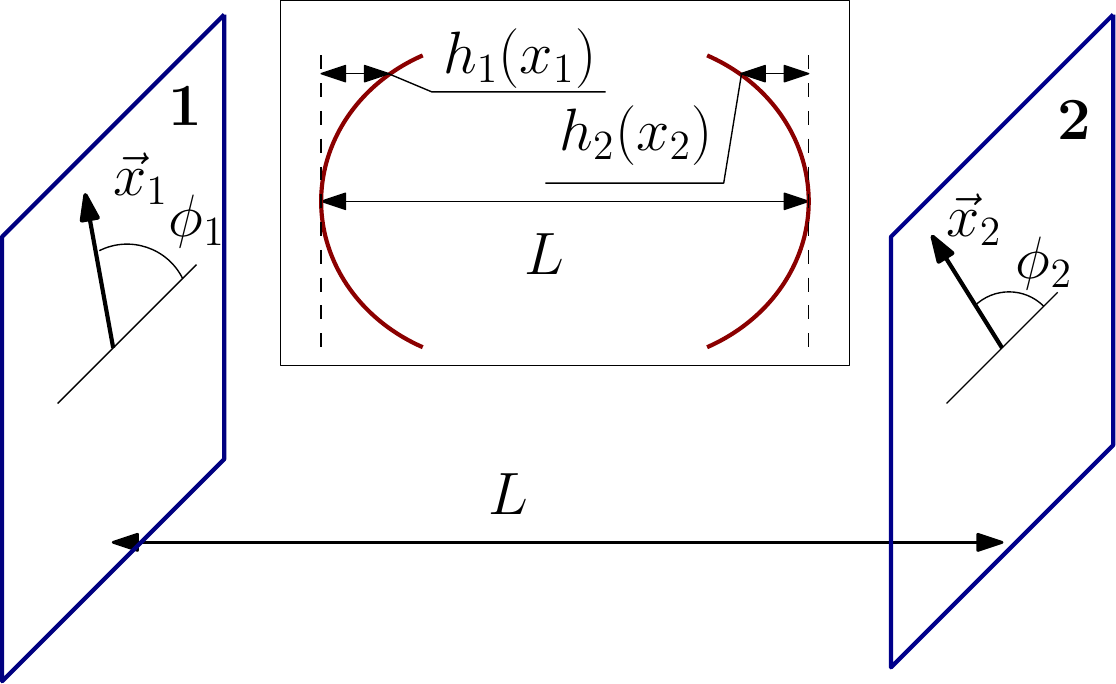}
 \caption{The field distribution $\Psi_2(\vec x_1)$ in plane 1 can be found from the distribution $\Psi_2(\vec x_2)$ in plane 2 using the Fresnel integral \eqref{Fresnel1}. Inset: FP cavity with axially symmetric mirrors, which shapes characterised by deviations $h_1(x_1),\ h_2(x_2)$.}\label{Fresnel}
\end{figure}

Let us consider a FP cavity consisting of two identical mirrors separated by distance $L$, as shown in Fig.~\ref{Fresnel}. For the sake of simplicity we introduce dimensionless variable $x$ and parameters $b,  a_m$ as follows:
\begin{align}
   \label{x}
 x & = \frac{r}{b},\quad b=\sqrt\frac{L}{k},\quad k= \frac{2\pi}{\lambda},
 \quad a_m=\frac{r_m}{b}\,,
\end{align}
where $r$ is the distance from the center of the mirror in the plane of the mirror (radial coordinate), $b$ is the scaling factor, $\lambda$ is the optical wavelength, $r_m$ is the radius of the mirror. The geometrical profile of the mirrors is described by dimensionless parameters
$h_{1,2}$ having meaning of a deviation of the mirror surface in the direction orthogonal to the mirror plane
\begin{align}
\label{h12}
 h_{1,2} &= k\, y_{1,2},
\end{align}
where $y_{1,2}$ is an actual physical (dimensional) deviation.
Selection of the mirror planes and distance between them has certain flexibility since the mirrors are not flat. We postulate the planes to be parallel. The distance $L$ between the planes is large enough ($L\gg r_m$) to apply the paraxial approximation.

Fresnel diffraction theory allows one to find a distribution of the electric field of an electromagnetic wave at any point of space if distribution of the field is known in a plane. For instance, the field distribution
$\Phi_2(\vec x_2)$ in plane $2$ (see Fig.~\ref{Fresnel}) defines distribution $\Phi_1(\vec x_1)$ in plane $1$. It can be evaluated via Fresnel integral with kernel $G(\vec x_1, \vec x_2)$ presented in dimensionless form as follows
\begin{subequations}
   \label{Fresnel1}
   \begin{align}
	\Phi_1(\vec{x_1}) &= \int G(\vec{x_1}, \vec{x_2}) \Phi_2(\vec{x_2}) d\vec{x_2} \\
	G(\vec{x_1}, \vec{x_2}) &= -\frac{i}{2\pi}\exp\left(i\left[\frac{|\vec{x_1}-\vec{x_2}|^2}{2}\right]\right)
   \end{align}
 \end{subequations}
where $\vec{x_1}$ and $\vec{x_2}$ are dimensionless radius vectors on planes 1 and 2,  shown in Fig.~\ref{Fresnel}, and the integration is performed over plane 2.

\subsection{Fourier and Hankel Transforms}

Since Eq.~\eqref{Fresnel1} is a convolution, the following formulas are valid for the Fourier transforms  $\tilde \Phi_{1,2}$ of distributions $\Phi_{1,2}$
\begin{align}
  \label{FresnelFT}
 \tilde \Phi_{1} (p,q) &= \tilde G(p,q) \, \tilde \Phi_2(p,q),\\
 \label{PhiFP}
 \tilde \Phi_{1,2} (p,q) &= \frac{1}{2\pi}\int \Phi_{1,2}(x_c,y_c) \, e^{-ipx_c-ipy_c}\, dx_c\, dy_c,\\
 \label{tildeG}
 \tilde G (p,q) &= \frac{-i}{2\pi} \exp\left(-\frac{i\rho^2}{2}\right), \quad \rho=\sqrt{p^2+q^2}.
\end{align}
Here parameters $x_c,\ y_c$ represent Cartesian coordinates in planes 1 and 2. We use Fourier transform in a ``symmetric'' representation \eqref{PhiFP} with the normalization factor $1/(2\pi)$ in front of the both the direct and the inverse transforms (not $1/(2\pi)^2$ in front of the direct transform only, as in the traditional form). We also assume that the mode field distributions have axial symmetry of order $\ell$, so the field amplitudes can be presented as
\begin{align}
\label{ellSym}
 \Phi_{1,2} (x_c,y_c) &= \Phi_{1,2}^{(\ell)}(|x_{1,2}|) \, e^{i\ell\phi_{1,2}},
\end{align}
where $|x_{1,2}|,\ \phi_{1,2}$ are polar coordinates. In this case the Fourier transform may be simplified to the Hankel transform
\begin{align}
  \tilde \Phi_{1,2} (p,q) &= (-i)^\ell e^{i\ell\theta}\,  \overline\Phi_{1,2}^{(\ell)}(\rho),\\
  \label{HTI}
  \overline \Phi_{1,2}^{(\ell)}(\rho) &= \int_0^\infty J_\ell(\rho x)\,\Phi_{1,2}^{(\ell)}(x)\, x\, dx
    =\mathbb  H_\ell\Phi_{1,2}^{(\ell)},
\end{align}
where $x\equiv |x_{1,2}|$ stands for the spatial dimensionless radial coordinate,  $J_\ell $ is the Bessel function of the first kind of order $\ell$. As shown above, it is convenient to introduce an operator $\mathbb H_\ell$ to write the Hankel transform  $\overline \Phi_{1,2}^{(\ell)}(\rho)$   of functions $\Phi_{1,2}^{(\ell)}(x)$.

The expression for the inverse Hankel Transform can be represented in a similar way as
\begin{align}
 \label{HTinvI}
 \Phi_{1,2}^{(\ell)}(x) &= \int_0^\infty J_\ell(\rho x)\,\overline \Phi_{1,2}^{(\ell)}(\rho)\, \rho\, d\rho
    =\mathbb  H_\ell^{-1}\overline \Phi_{1,2}^{(\ell)}
\end{align}

Using the operator notations we rewrite Eq.~\eqref{Fresnel1} for the radial functions $\Phi_{1,2}$ in a short from
\begin{align}
  \label{PhiG}
  \Phi_1^{(\ell)} &=  \mathbb P_\text{plane} \Phi_2^{(\ell)},\quad  \mathbb P_\text{plane}= \mathbb  H_\ell^{-1}\tilde G \mathbb H_\ell,
\end{align}
where $\mathbb P_\text{plane},\ \mathbb H_\ell,\ \mathbb H_\ell^{-1}$ are the integral operators and $\tilde G $ is a function.

\subsection{Axial symmetric mirrors}\label{ASmirrors}

We consider a FP cavity represented by curved axial symmetric mirrors $1$ and $2$ defined by dimensionless parameters $h_1(x_1)$ and $h_2(x_2)$ \eqref{h12}, where $x_{1,2} \equiv |\vec x_{1,2}|$. For a particular case of spherical mirrors with curvature radii $R_{c1,c2}$ these parameters are equal to
\begin{align}
 h_{1,2}^\text{sph}= \frac{x_{1,2}^2}{2\rho_{1,2}}, \quad \rho_{1,2}= \frac{R_{c1,c2}}{L}.
\end{align}
We obtain for the paraxial approximation of the field distributions  $\Psi_{1,2}$ of the cavity modes of order $\ell$
\begin{subequations}
\label{PsiPhi}
\begin{align}
 \Psi_{1,2}(\vec x_{1,2}) &= \Psi_{1,2}^{(\ell)}(x_{1,2})\, e^{i\ell \theta} ,\\
 \Psi_{1}^{(\ell)}(x_{1}) &= \Phi_1^{(\ell)}(x_1) \, e^{-ih_1},\\
 \Psi_2^{(\ell)}(x_2) &= \Phi_2^{(\ell)}(x_2) \, e^{i h_2},
\end{align}
\end{subequations}
where $\Phi_{1,2}$ are the field distributions in planes $1$ and $2$, and $\Psi_{1,2}$ are the field distributions on surfaces of the mirrors.

Using \eqref{PhiG} we present radial distribution $\Psi^{(\ell)}_{1}(\vec x_{1})$ via $\Psi_{2}^{(\ell)}(\vec x_{2})$ in the operator form
\begin{align}
  \label{PsiG}
  \Psi_1^{(\ell)} &= \mathbb P_\text{forward}^{2\to 1}\, \Psi_2^{(\ell)},\\
  \label{PforwardI}
  \mathbb P_\text{forward}^{2\to 1}&=
    \left(\mathbb R_1 \mathbb  H_\ell^{-1}\tilde G \mathbb H_\ell \mathbb R_2\right),\quad
    \mathbb R_{1,2} = e^{-ih_{1,2}}
\end{align}
 The same relation can be written in a form of integral equation
\begin{subequations}
   \label{Fresnel2}
   \begin{align}
	\Psi_1^{(\ell)}(x_1) &= \int g(x_1, x_2)\, \Psi_2^{(\ell)}(x_2)\, x_2\, dx_2 \\
	g(x_1, x_2) &= - i^{\ell+1} J_\ell(x_1x_2) \times\\
	& \times \exp\left( i\left[\frac{x_1^2+x_2^2}{2}\right] \nonumber
	 - i h_1(x_1)- ih_2(x_2)\right).
   \end{align}
\end{subequations}

In the case of the identical mirrors we assume that the spatial profile of the field does not change over the round trip: $\Psi_1(x)=\Lambda \Psi_2(x)$ (where $\Lambda$ is a complex number) and introduce $\Psi(x)\equiv \Psi_2(x)$. To find the spatial field distributions of an eigenmode of the cavity we have to solve an eigenvalue problem
\begin{align}
   \label{initLambda}
	\Lambda\Psi(x_1) &= \int g(x_1, x_2)\, \Psi(x_2)\, x_2\, dx_2.
\end{align}
The integral in \eqref{initLambda} is taken over the mirror surface, $\Lambda$ is the eigenvalue that contains information both about the frequency and corresponding attenuation of the cavity mode of interest, $\Psi(x)$ is the eigenfunction showing the spatial distribution of the field at the mirror surface.

Similarly, for the case of non-identical mirror we have to find the modification of the field distribution after the {\em round trip} propagation of a wave in the cavity and postulate that the field distribution does not change. The eigenvalue problem can be written in the operator form
\begin{align}
   \label{initLambda2}
	\Lambda\Psi^{(\ell)} &= \mathbb P_\text{round}\, \Psi^{(\ell)},\\
	\mathbb P_\text{round}& = \mathbb P_\text{forward}^{1\to 2}\cdot \mathbb P_\text{forward}^{2\to 1}=\\
	&=\left(\mathbb R_2 \mathbb  H_\ell^{-1}\tilde G \mathbb H_\ell \mathbb R_1\right)
	\left(\mathbb R_1 \mathbb  H_\ell^{-1}\tilde G \mathbb H_\ell \mathbb R_2\right)
\end{align}
We see that matrices $\mathbb R_{1,2}$, accounting mirror's profile, are included twice.

\begin{table}[ht]
\caption{Parameters of the aLIGO FP cavity}\label{param}
\begin{tabular}{|l|c|}
\hline
Parameter & Value \\
\hline
Arm length, $L$ & $4$~km\\
Optical wavelength, $\lambda$ & $1064$~nm\\
Intracavity power, $P$ & $800$~kW\\
$AS_{00}$ mode round trip loss, ${\mathcal L}$ & 0.41~ppm\\
$D_{10}$ mode round trip loss, ${\mathcal L}$ & 10~ppm\\
Characteristic cavity length $b=\sqrt{L\lambda/2\pi}$ & $0.0260$~m\\
Radius of mirrors, $R$ &$0.17$~m\\
Dimensionless mirror radius $a_m=R/b$ & $6.53$ \\
Radius $w$ of laser spot at the mirror & $0.06$~m \\
Radius $w_0$ of laser beam at the waist  & $ 0.0115$~m\\
Curvature radius of spherical mirrors, $R_c$ & $2076$~m\\
Geometric parameter $g=1-L/R_c$  of the cavity & $-0.92649$\\
Gouy phase, $\arctan \left[ (b/w_0)^2\right ]$ & 1.378\\
\hline
\end{tabular}
\end{table}

\subsection{Numerical values}

We utilize original aLIGO FP cavity parameters in our numerical simulations and assume a symmetric case using the same radii of curvatures for ITM and ETM. The parameters are listed in Table~\ref{param}. In this article we are interested primarily in the fundamental attenuation and study diffraction loss of the main mode of a LIGO interferometer. It is known to be 0.45~ppm corresponding to a FP cavity with perfect spherical mirrors (without roughness or tilt).
%

It will be worth noting, though, that the round trip loss of an aLIGO interferometer is measured to be around 100ppm. The round trip loss calculated using coated mirror surface maps and large angle scattering measurements can account only 40 ppm. Further study is needed to understand this discrepancy.

\section{Gaussian approximation of FP cavity modes}

Gaussian beams \cite{Siegman1986, Saleh1991, Marcuse1972, Svelto2010} represent exact solutions of \eqref{initLambda} for a FP cavity with spherical mirrors which are indefinitely large in transversal direction. The "infinite" spherical mirrors do not have any practical sense and Gaussian modes can be approximately applied for a finite sized spherical mirror if its  radius, $r_m$, is much larger than radius $w$ of the beam spot on the mirror. The dependence of amplitude of the electromagnetic field on the distance $r$ from the beam center is $\sim e^{-r^2/w^2}$, so it is reasonable to assume that the mirror is infinite if $r/w>3$. In what follows we consider the case of high finesse cavity and derive an analytical expression useful for an approximate evaluation of the optical loss due to the finite mirror size.

\subsection{Cavity losses in the clip approximation}

To find the loss of a high finesse cavity we first solve \eqref{initLambda} and find the eigenfunction of the cavity field in the lossless approximation. The round trip diffraction loss of each mode with known field distribution $\Psi(x)$ at the mirror surface can be estimated then by the so called {\em clip} approximation. In this approximation we evaluate the part of the field of the eigenmode that is not confined by the mirror and consider it as a loss per single reflection. Mathematically it means that the power loss per round trip is
\begin{align}
   \label{clip}
 \mathcal L_\text{clip} = 2 \, \frac{\int_{a_m}^\infty|\Psi(x)|^2 x\, dx}{\int_0^\infty|\Psi(x)|^2 x\, dx},
\end{align}
where $\Psi(x)$ is one of Laguerre-Gaussian solution of \eqref{initLambda} for FP cavity with {\em infinite} large spherical mirrors, and factor $2$ comes from the presence of two mirrors in the cavity.

For the parameters of a FP cavity listed in Table~\ref{param} the diffraction loss of the main Gaussian mode is
\begin{align}
   \label{clip2}
 \mathcal L_0 = 2 \exp\left( - \frac{2 r_m^2}{w^2}\right) \simeq 0.21\ \text{ppm}.
\end{align}
where $ppm$ stands for the part per million. We have found that the exact numerical simulation shows that the mode has approximately twice larger diffraction loss (see Tables~\ref{param} and Table \ref{table1} below). It means that the clip approximation is suitable rather for a qualitative, not quantitative, analysis.

\subsection{Loss due to mirror tilt in the clip approximation}

The clip approximation is useful to understand loss occurring in a FP resonator with tilted mirrors. The tilt results in the geometrical mismatch between the pump light and the resonator as well as in the decrease of the finesse and quality factor of the resonator. In this section we analyze both effects.

The resonant steady state amplitude $B$ of the field inside an ideal FP cavity can be expressed through the amplitude $A$ of the incident wave as (for example, see \cite{Siegman1986})
\begin{align} \label{binside}
 B &= \frac{2\sqrt T}{T+\mathcal L_0}\cdot A,
\end{align}
where $T$ is power transmittance of input mirror (IM),  end mirror (EM) is assumed to be perfectly reflecting, $\mathcal L_0$ is round trip diffraction loss, which may be calculated in the clip approximation \eqref{clip}.
\begin{figure}[b]
 \includegraphics[width=0.45\textwidth]{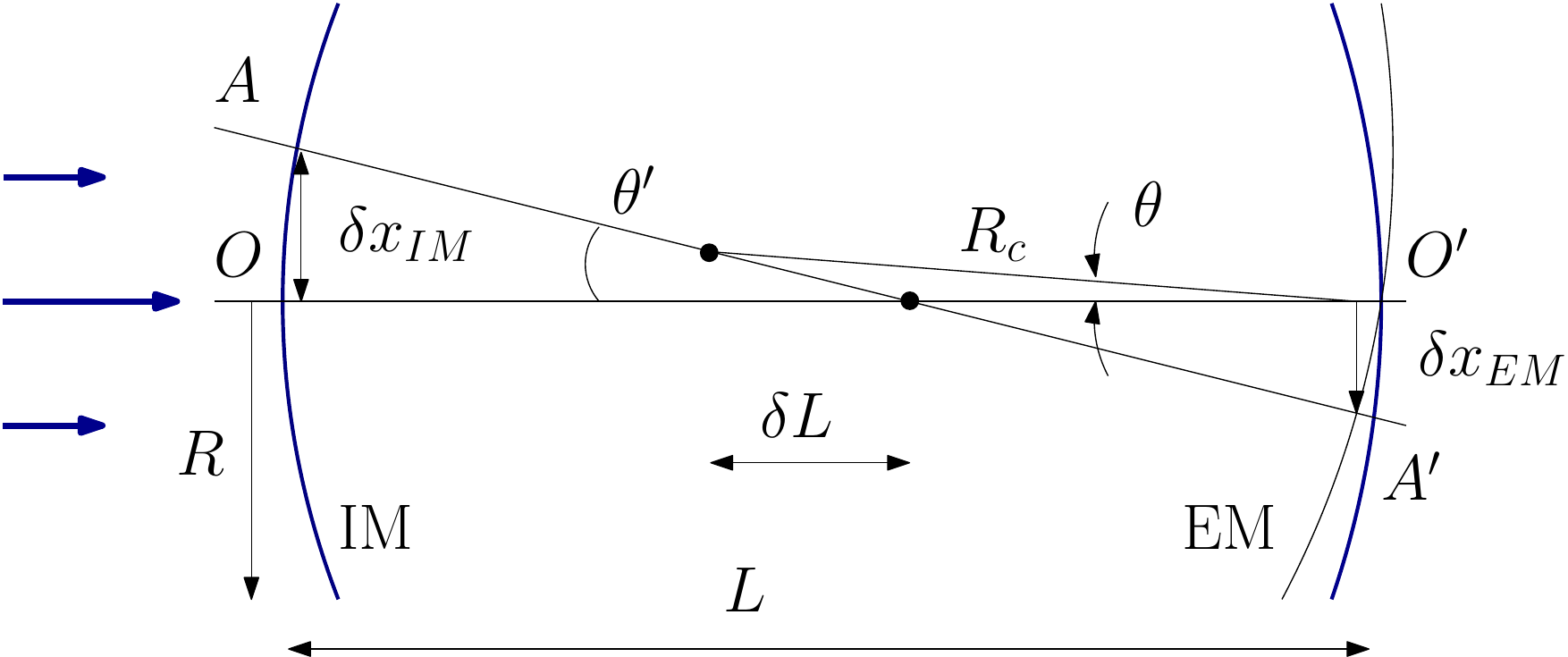}
 \caption{A Fabry-Perot cavity with two identical spherical mirrors with curvature radii $R_c$, $\delta L=2R_c-L$. Radii of mirrors are $R$. The end mirror (EM) is tilted by small angle $\theta$. As the result the axis of the cavity shifts from OO' to AA'. The mode center becomes closer to the mirror edge and the loss due to diffraction of the beam increases. }\label{geomYa}
\end{figure}

Let us consider a FP cavity with two identical spherical mirrors one of which (for instance, the end mirror, EM) is tilted by a small angle $\theta$, as shown on Fig.~\ref{geomYa}.  A small tilt produces change of optical axis position (AA' instead of QQ') while field distributions on mirrors are shifted:
\begin{align}
  \tilde \psi_B^\text{IM} & = \Psi_B(x - \delta x_\text{IM},y),\\
  \tilde \psi_B^\text{EM} & = \Psi_B(x + \delta x_\text{EM},y)
\end{align}
Here and below we denote by tilde distributions for cavity with tilted EM.

We present the amplitude of the field $\tilde B$ inside the cavity with tilted EM (in resonance) in a way similar to Eq.~(\ref{binside}):
\begin{align}
\label{tildeB}
 \tilde B & = \frac{2\sqrt T\, \tilde A}{T+ \tilde{\mathcal L}_0},
\end{align}
to take into account two reasons of decrease of amplitude $\tilde B$ inside tilted cavity comprising decrease of the effective pump amplitude due to mode mismatch for the cavity and the external pump as well as increase of the diffraction loss. The decrease of the effective pump  amplitude due to the mode mismatching occurs because the  field distribution of the pump light does not coincide with the shifted distribution of the cavity mode. As a result, the input amplitude $A$ in the formula (\ref{binside}) should be replaced by {\em smaller} one $\tilde A$ \cite{Hiro2015}:
  \begin{align}
  \label{A1}
  \tilde A &= A\cdot \int\Psi_A\tilde \Psi_B\, d\vec r
   \simeq  A\left(1 -\frac{\theta^2}{\theta_A^2}\right),\\
   \label{thetaA}
   \theta_A &\simeq \frac{\sqrt 2\, w\,\delta L }{R_c^2},\quad
   \int|\tilde \Psi_B|^2\, d\vec r \simeq 1\,.
\end{align}

The increase of diffraction loss $\tilde {\mathcal L}_0$ results from the reduction of the spatial overlap between the cavity mode and the mirror. The modified loss value can be estimated by the clip approximation \cite{Hiro2015}:
 \begin{align}
  \label{tloss}
  \tilde {\mathcal L}_0 &= \mathcal L_0\left(1 + \frac{\theta^2}{\theta_\text{perm}^2}\right),\\
      \theta_\text{perm}&\simeq \frac{\delta L w^2}{ \sqrt 2\, R_c R\sqrt{R_c^2+(L-R_c)^2}}
      \simeq \frac{\delta L w^2}{ 2 R_c^2 R}.
\end{align}
Here we accounted that a) $\delta_{xIM} = R_c^2 / \delta{L} \times \theta_{ETM}$ and $\delta_{xEM} = R_c (L-R_c) / \delta{L} \times \theta_{ETM}$, b) the round trip loss is proportional to $\delta_{xIM}^2 + \delta_{xEM}^2$, c) for aLIGO parameters $L-R_c\simeq R_c$. Here we introduce $\theta_\text{perm}$ so that at $\theta= \theta_\text{perm}$ round trip loss increases by 2 times.

Substituting (\ref{thetaA}, \ref{tloss}) into  \eqref{tildeB} we obtain an expression for the decrease of the intracavity amplitude when the mirror is tilted
\begin{align}
\label{tildeB2}
 \tilde B &\simeq  \frac{2\sqrt T\,A}{T+\mathcal L_0}\cdot
  \left(1 - \frac{\theta^2}{\theta_\text{eff}^2} \right),\\
  \label{thetaeff}
  &\frac{1}{\theta_\text{eff}^2}  = \frac{1}{\theta_A^2} + \frac{\mathcal L_0}{T+\mathcal L_0}\cdot \frac{1}{\theta_\text{perm}^2}.
\end{align}
This field amplitude decrease will be observed in the pumped cavity in the steady state. The decrease of the finesse of the cavity, though, is defined by the $\theta_\text{perm}$ which can be estimated for the LIGO FP cavity (see Table~\ref{param} for the parameters) as
\begin{equation}
 \label{thetaGeom}
 \theta_\text{perm}^\text{geom}\simeq 0.37 \cdot 10^{-6}
\end{equation}
Presented here estimation is an approximate one and differs by a few times from the results of a more accurate successive approximation technique (see the detailed formulas in Sec.~\ref{SmallTilt} and in Appendix~\ref{app1}). The difference can be explained by the sensitivity of the field distribution inside of the cavity on the mirror tilt. The derivation of the formulas (\ref{thetaA}, \ref{tloss}) is based on the assumption of the tilt-insensitive Gaussian distribution as well as known position of the mode axis shift in the cavity with the tilted mirror. The successive approximation technique takes the change of the eigenmode geometrical profile into account and the resultant loss becomes more pronounced. For example, for the numerical parameters listed in Table~\ref{param} we obtain an estimate

\begin{align}
 \label{thetaSAG}
 \theta_\text{perm}^\text{saG} \simeq 1.1\cdot 10^{-6}.
\end{align}
We see that the loss values \eqref{thetaGeom} and \eqref{thetaSAG} differ by about 3 times. The difference becomes even larger for the case of a cavity with non-spherical mirrors. The main reason that the clipping approximation underestimates the actual loss is due to the Airy pattern like tail induced by the finite aperture of the mirrors.  For example, even if initially beam on mirror 2 has Gaussian shape, the field on mirror 1 coming from mirror 2  is not a Gaussian, but it has a long non Gaussian tail induced by the finite size mirror 2.

\section{Numeric Simulation of a FP cavity}

To perform the numerical simulations taking into account diffraction loss we assume that the field distribution of a mode is limited by a circle of dimensionless radius $a$ which is larger than the cavity mirror radius $a_m$ \eqref{x}. We introduce window parameter $S$ as
\begin{align}
  \label{S}
   S=\frac{a}{a_m}  >1.                                                                                                                                                                                                                                                                                                                                                                                                                                                                                                                                                                                                                                                                                                         \end{align}
Optimal selection of the free parameter $S$ is discussed in Section \ref{window}.

\subsection{Orthogonal basis}

We consider the axial symmetry modes of order $\ell$ \eqref{ellSym} and find a complete orthogonal basis of functions $\varphi^{(\ell)}_{k}(x)$ in order to evaluate radial distributions $\Psi^{(\ell)}_{1,2}(x)$ of the modes. We select this basis in form
\begin{align}
 \label{phi}
   \varphi^{(\ell)}_{k}(x) = J_{\ell}\left(\xi_{k}\frac{x}{a}\right),
\end{align}
where $k$ is the index of the function (a natural number). The coefficients $\xi_k$ are selected in a way to achieve the basis orthogonality. We define scalar product as
\begin{align}
   \label{scalarP}
	&\left\langle\varphi^{(\ell)}_{k}(x), \varphi^{(\ell)}_{n}(x)\right\rangle
	 = 2\pi \int\limits_0^a \varphi^{(\ell)}_{k}(x)\, \varphi^{(\ell)}_{n}(x)\, x\, dx=\\
	& = \frac{a^2}{\xi_n^2 -\xi_k^2}\Big[\xi_nJ_{\ell+1}(\xi_n) J_{\ell}(\xi_k) - \xi_k J_{\ell+1}(\xi_k) J_{\ell}(\xi_n)\Big]
\end{align}
Here  formula (1.8.3.10) from \cite{83PrudnikovS} was utilized to derive this expression.
We require the basis to be orthogonal
\begin{align}
 \left\langle\varphi^{(\ell)}_{k}(r), \varphi^{(\ell)}_{n}(r)\right\rangle &= 0,\quad \text{if}\ n\ne k
\end{align}
This condition is fulfilled if coefficient $\xi_{n}$ is a solution of equation
\begin{equation}
\label{root_eq}
P J_{\ell}(x) - Q\, x J_{\ell+1}(x) = 0,
\end{equation}
where $P$, $Q$ are arbitrary numbers (c.f. the analogue presented for $\ell=0$ in \cite{93vinet}).

The decomposition is possible according to the Steklov theorem applied to the Sturm-Liouville problem. The orthogonal basis has unlimited number of orthogonal functions. For the sake of simplicity we use a finite set of basis modes in our simulations $\{\xi_k\}_{k=1}^N$ with $N = 512$ or $N=1024$. Dependence of calculated numerically diffraction losses from $N$ is presented below on Fig.~\ref{L(N)}. It is easy to see that the result of simulation does not change more than 10\% if $N>500$.

\subsection{Normalization}

It is convenient to normalize the orthogonal basis we have selected. One may calculate
\begin{align}	\left\langle\left(\varphi^{(\ell)}_{n}(r)\right)^2\right\rangle = 2\pi \int\limits_0^a J_{\ell}^2\Big(\xi_n\frac{r}{a}\Big)rdr \equiv \pi a^2 \mathcal{N}_n^{(\ell)},
\end{align}
where $\mathcal{N}_n^{(\ell)}$ is a dimensionless parameter defined as
\begin{align}
\label{Nell}
\mathcal{N}_n^{(\ell)} = \left\{ \begin{array}{ll} J_{\ell}^2(\xi_n)\Big(1+\frac{P}{Q\xi_n^2}\Big[\frac{P}{Q}-2\ell\Big]\Big) & \text{if} \quad Q \neq 0, \\
	J_{\ell+1}^2(\xi_n) & \text{if} \quad Q = 0
\end{array} \right.
\end{align}
for arbitrary $P$ and $Q$. The expression (1.8.3.12) \cite{83PrudnikovS} was utilized to perform the analytical integration.

The Hankel Transform function $\overline \Psi^{(\ell)}_{1,2}(\rho)$ can be decomposed using the orthogonal basis in a similar way
\begin{align}
\label{psiell}
 \psi^{(\ell)}_{k}(\rho) = J_{\ell}\left(\xi_{k}\frac{\rho}{b}\right)
\end{align}
defined in a finite circle of radius $b$.

The scalar product of basis functions $\psi^{(\ell)}_k$ can be presented as
\begin{align}
 \label{scalarP2}
	&\left\langle\psi^{(\ell)}_{k}(\rho), \psi^{(\ell)}_{n}(\rho)\right\rangle =\\
	&\qquad = 2\pi \int\limits_0^b J_{\ell}\left(\xi_{k}\frac{\rho}{b}\right)J_{\ell}\left(\xi_{n}\frac{\rho}{b}\right) \rho d\rho
	 = \pi b^2 \mathcal N_n^{(\ell)}\, \delta_{nk}\,.\nonumber
\end{align}
Here constant $ \mathcal N_n^{(\ell)}$ is defined by the same formula \eqref{Nell}.

Therefore, we can present any function $\Psi^{(\ell)}(x)$ and its Hankel Transform $\overline \Psi^{(\ell)}(\rho)$ as expansion in series over the introduced orthonormal basis
\begin{align}
	\Psi^{(\ell)}(x) &= \sum_{k=1}^{\infty} c_k \varphi^{(\ell)}_k(x), \quad
	\label{series}
	\overline\Psi^{(\ell)} (\rho) = \sum_{k=1}^{\infty} d_k \psi^{(\ell)}_k(\rho)
\end{align}
where $c_k$ and $d_k$ are decomposition coefficients.

\subsection{Discrete Hankel Transform}

We consider  function $\Psi^{(\ell)}(x)$ with axial symmetry of order $\ell$ defined in a circle with radius $a$ and its Hankel Transform $\overline\Psi^{(\ell)} (\rho)$ defined in a circle with radius $b$
\begin{align}
\label{HT2}
 \Psi^{(\ell)}(x) &= \int_0^b J_\ell(\rho x)\,\overline\Psi^{(\ell)} (\rho)\, \rho\, d\rho
\end{align}
This is a {\em finite-sized} Hankel Transform, so the integration limit is finite, unlike the one in Eq.~\eqref{HTI}.

By substituting \eqref{series} into \eqref{HT2} we obtain
\begin{align}
   \label{Psiellx}
 \Psi^{(\ell)}(x) &=\sum_{k=1}^{\infty} d_k \int_0^b J_\ell(\rho x)\, J_{\ell}\left(\xi_{k}\frac{\rho}{b}\right)\, \rho\, d\rho	
\end{align}
Selecting sampling points $x=\xi_k/b$ in \eqref{Psiellx} and using orthogonality \eqref{scalarP2} we find
\begin{align}
   \label{Psiell2}
 \Psi^{(\ell)}\left(\frac{\xi_k}{b} \right) = d_k\, \frac{b^2 \mathcal N_k^{(\ell)}}{2} \,.
\end{align}
Hence, expressing $d_k$ from \eqref{Psiell2} we can rewrite \eqref{series} in form
\begin{align}
 \overline \Psi^{(\ell)}(\rho) &= \sum_{k=1}^{\infty} \frac{2}{b^2 \mathcal N_k^{(\ell)}}\, \Psi^{(\ell)}\left(\frac{\xi_k}{b} \right)
      J_{\ell}\left(\xi_{k}\frac{\rho}{b}\right)
\end{align}

Finally, selecting $\rho= \xi_\alpha/a$ we obtain
\begin{align}
 \overline \Psi^{(\ell)}\left(\frac{\xi_\alpha}{a}\right) &= \sum_{k=1}^{\infty} \frac{2}{b^2 \mathcal N_k^{(\ell)}}\,
   \Psi^{(\ell)}\left(\frac{\xi_k}{b} \right)J_{\ell}\left(\frac{\xi_k\xi_\alpha}{ab}\right)
\end{align}
This is a {\em discrete} Hankel Transform of order $\ell$ presented as a discrete linear operation acting on vector $\Psi^{(\ell)}\left(\frac{\xi_k}{b} \right)$ and giving output vector $\overline \Psi^{(\ell)}\left(\frac{\xi_\alpha}{a}\right)$ expressed as a matrix product
\begin{align}
   \label{HTdir}
 \overline \Psi^{(\ell)}\left(\frac{\xi_\alpha}{a}\right) &= \sum_{k=1}^{\infty} \mathbf H_{\alpha k}^{(\ell, +)}\,
    \Psi^{(\ell)}\left(\frac{\xi_k}{b}\right) \\
   \mathbf H_{\alpha k}^{(\ell, +)} &=
   \frac{2}{b^2 \mathcal N_k^{(\ell)}}\,
  J_{\ell}\left(\frac{\xi_k\xi_\alpha}{ab}\right)
\end{align}
Formula for reciprocal discrete Hankel Transform can be derived in a similar way:
\begin{align}
   \label{HTinv}
  \Psi^{(\ell)}\left(\frac{\xi_k}{b}\right) &= \sum_{\alpha=1}^{\infty} \mathbf H_{\alpha k}^{(\ell, -)}\,
   \hat \Psi^{(\ell)}\left(\frac{\xi_\alpha}{a}\right) \\
   \mathbf H_{\alpha k}^{(\ell, -)} &=
   \frac{2}{a^2 \mathcal N_\alpha^{(\ell)}}\,
   J_{\ell}\left(\frac{\xi_k\xi_\alpha}{ab}\right)
\end{align}

\subsection{A finite basis selection}

Let us discuss several important properties of the discrete Hankel Transform introduced above and select a basis convenient for the numerical simulations. The number of discrete points $r_k/b$ ($\rho_k/a$) in the direct (Hankel) space is assumed unlimited. The wave functions are defined inside of finite circles of radii $a$ and $b$. It leads to the restriction
\begin{align}
 \frac{\xi_k}{b} \leq a,\quad \frac{\xi_\alpha}{a} \leq b\ \Rightarrow\ \xi_k \leq ab
\end{align}
It is convenient to select upper index $N$ and window radius $a$ in the direct space so that
\begin{align}
 b=\frac{\xi_N}{a}\, .
\end{align}
Then discrete points $x_k$ in the direct space and the points $\rho_\alpha$ in Hankel space can be selected as
\begin{align}
   \label{xk}
 x_k &= \frac{\xi_k}{\xi_N}\, a ,\quad x_N=a,\\
 \label{rhoalpha}
 \rho_\alpha &=\frac{\xi_\alpha}{a}, \quad \rho_N= b=\frac{\xi_N}{a}
\end{align}

In this case the the infinite sums in (\ref{HTdir}, \ref{HTinv}) should be replaced with finite ones ($1\dots N$) and  matrices $\mathbf H^{(\ell, \pm)}$ become
\begin{align}
  \mathbf H_{\alpha k}^{(\ell, +)} &=    \frac{2 a^2}{\xi_N^2 \mathcal N_k^{(\ell)}}\,
   J_{\ell}\left(\frac{\xi_k\xi_\alpha}{\xi_N}\right),\\
   \mathbf H_{\alpha k}^{(\ell, -)} &=
   \frac{2}{a^2 \mathcal N_\alpha^{(\ell)}}\,
   J_{\ell}\left(\frac{\xi_k\xi_\alpha}{\xi_N}\right)
\end{align}

In general case the truncated matrixes are inconvenient for numeric simulations since
 \begin{align}\textbf{H}^{(\ell, +)}\times\textbf{H}^{(\ell, -)} \neq \textbf{I}
 \neq \textbf{H}^{(\ell, -)}\times\textbf{H}^{(\ell, +)}.
 \end{align}
This property leads to a divergence of the iterative calculations. Instead we utilize matrices $\textbf{H}^{(\ell, +)}$ and $\left(\textbf{H}^{(\ell, +)}\right)^{-1}$ in the direct and reciprocal Hankel Transform. It is also possible to select the complimentary ($\textbf{H}^{(\ell, -)}$ and $\left(\textbf{H}^{(\ell, -)}\right)^{-1}$) matrix pair. The relative difference between the solutions found in these two ways for the fundamental mode (AS00) of a lossless FP cavity does not exceed $10^{-13}$ for $N=512$.

The finite discrete Hankel Transform operators  $\textbf{H}^{(\ell, +)}$ and $\left(\textbf{H}^{(\ell, +)}\right)^{-1}$ (or $\textbf{H}^{(\ell, -)}$ and $\left(\textbf{H}^{(\ell, -)}\right)^{-1}$) represent the integral Hankel Transform operators $\mathbb H_\ell,\ \mathbb H_\ell^{-1}$ (\ref{HTI}, \ref{HTinvI}). We use the discrete Hankel Transform operators because they can be presented in the matrix form. It is convenient for numeric simulations.

 \subsection{Discrete propagator of evolution from mirror to mirror}

Using the convolution theorem we perform the same analysis as in Sec.~\ref{model} and obtain a discrete analogue of the integral operator $\mathbb P_\text{plane}$ \eqref{PhiG}
\begin{align}
  \textbf{P}_\text{plane}^{(\ell)} = \left(\textbf{H}^{(\ell, +)}\right)^{-1} \tilde{\textbf G}\mathbf{H}^{(\ell, +)}
\end{align}
where $\tilde{ \textbf G}$ is a Fourier Transform of Green function in the paraxial approximation (compare with $\tilde G$ in \eqref{tildeG} using \eqref{rhoalpha}):
\begin{align}
  \label{tildeGb}
	\tilde{\textbf G}_{\alpha\beta} = \exp\left(-\frac{i}{2}\cdot \frac{\xi_\alpha^2}{a^2}\right)\delta_{\alpha\beta}
\end{align}

The discrete propagator $\textbf{P}_\text{plane}^{(\ell)}$ describes evolution of the light beam propagating from a plane 1 to plane 2 (see Fig.~\ref{Fresnel}).

\subsection{Attenuation matrix}

In order to write the forward trip propagator $\mathbf P_\text{forward}^{(\ell)\, 2\to 1}$ for evolution of the light propagating from  mirror 2 to mirror 1 we repeat the procedure described in subsection \ref{ASmirrors}. As a result we obtain
\begin{align}
   \label{Psi2t01}
 \Psi^{(\ell)}_1 &= \mathbf P_\text{forward}^{(\ell)\, 2\to 1} \Psi_2^{(\ell)},\\
 \label{PforwardD}
  \mathbf P_\text{forward}^{(\ell)\, 2\to 1} &= \mathbf R_1 \mathbf P_\text{plane}^{(\ell)}\mathbf R_2.
\end{align}
Formula \eqref{Psi2t01} is suitable to evaluate the radial part $\Psi^{(\ell)}_1$ of field distribution on mirror 1 (presented as a finite column of numbers) using the (known) radial part $\Psi^{(\ell)}_2$ of the field distribution on mirror 2 through a matrix product. The assumption of the axial symmetry of order of $\ell$ allows reducing the 2D diffraction problem to a 1D one. This is an advantage of the proposed method.

The attenuation matrices $\mathbf R_{1,2}$ in \eqref{PforwardD} account for curvature, reflectivity and finite size of mirrors. For axial symmetric mirrors these matrices are diagonal
\begin{align}
   \label{R1}
 (\mathbf R_1)_{kn} = \exp\big[-ih_1(x_k)\big]\, D_k\,\delta_{kn}
\end{align}
Here the first multiplier is analogues to multiplier $\mathbb R_1$ in \eqref{PforwardI} (we assume perfectly reflected mirror), the coefficients $D_{k}$ represent the diaphragm function which sets radius $a_m$ of the mirror
\begin{equation}
\label{diaphragm}
 D_k =\left\{
   \begin{array}{cl}
    1,\quad & \text{if } x_k\le a_m,\\
    0,\quad & \text{if } x_k> a_m
   \end{array}
 \right.
\end{equation}
The radius $a_m$ is $S$ times less than the radius $a$ of the simulation area \eqref{S}. This is necessary to take the diffraction loss into account.

The round trip evolution of light in the FP cavity (from mirror 2 to mirror 1 and backward) is described by propagator
\begin{align}
 \label{ProundDb}
  \mathbf P_\text{round}^{(\ell)} &= \mathbf P_\text{forward}^{(\ell)\, 1\to 2}\mathbf P_\text{forward}^{(\ell)\, 2\to 1}=\\
  \label{ProundD}
  &=\left(\mathbf R_2 \mathbf P_\text{plane}^{(\ell)}\mathbf R_1\right)
   \left(\mathbf R_1 \mathbf P_\text{plane}^{(\ell)}\mathbf R_2\right).
\end{align}
The attenuation matrices $\bf R_1$ and $\bf R_2$ are included twice due to the fact that each matrix describes mirror geometrical profile which adds an additional phase. Formally, this is a consequence of the relation \eqref{ProundDb} between the round trip and forward trip propagators (compare with \eqref{initLambda2}).

Formula \eqref{ProundD} can be generalized to account reflection loss of each mirror by formal modification of formula \eqref{R1}:
\begin{align}
   \label{R1b}
 (\tilde{\mathbf R}_1)_{kn} = \exp\big[-ih_1(x_k)\big]\, D_k\,\sqrt{R_k}\,\delta_{kn}
\end{align}
where $R_k$ is {\em amplitude} coefficient of refraction of mirror depending on radial coordinate.

Expression (\ref{ProundD}) may be simplified if the mirrors are identical ($\bf R_1 = R_2 \equiv R$)
\begin{equation}
	\mathbf P_\text{round}^{(\ell)} = \left(\mathbf R \mathbf P_\text{plane}^{(\ell)}\mathbf R\right)^2
\end{equation}

\subsection{Eigenvalue problem in the discrete basis}

Propagation of light in a FP cavity is accompanied by diffraction loss due to a finite size of the mirrors and optical attenuation in the mirrors. It means that the overall amplitude of the light beam decreases from a round trip to a round trip. We assume that the spatial profile of the eigenmodes of the cavity does not change  and set the eigenvalue problem to find it
\begin{equation}
	\label{eigenvalue}
	\Lambda \Psi^{(\ell)} = \mathbf P_\text{round}^{(\ell)}\Psi^{(\ell)}, \quad \quad 	\mathcal{L} = 1 - |\Lambda|^2
\end{equation}
where $\mathcal{L}$ is round trip  loss and $\Psi^{(\ell)}$ is an eigenmode of the optical cavity. In case of the ideal mirrors (zero optical loss in the mirror coating) only diffraction loss remains valid. The propagator $\mathbf P_\text{round}^{(\ell)}$ depends on the azimuthal number $\ell$ of the mode.

Therefore, in this section we introduced a framework for a numerical analysis of the eigenvalue problem of a FP cavity. The approach is suitable for a FP cavity assembled with mirrors having {\em arbitrary axially symmetric} profile. The propagator $ \mathbf P_\text{round}^{(\ell)}$ is a finite matrix. The original infinite space eigenvalue problem reduces to a finite eigenvalue problem for matrix $ \mathbf P_\text{round}^{(\ell)}$.

\subsection{Accuracy of the numerical approach}

To evaluate the accuracy of the numerical simulation performed using the described above approach we i) compare the result of the numerical simulation with the result of the analytical approximation and ii) evaluate the stability of the numerical scheme by varying the parameters of the simulation.

The evaluation of the method accuracy using the results of the analytical calculations is not very accurate. An analytical solution in paraxial approximation (Laguerre-Gaussian beams) is valid only for a cavity assembled by a very large (strictly speaking, infinite) {\em spherical} mirrors. For a radius of the mirror larger than the beam radius one can {\em approximately} use a Laguerre-Gauss beam approximation for the eigenfunctions and estimate the diffraction loss in the clip approximation \eqref{clip}. The analytical clip approximation underestimates the actual loss due to the Airy pattern-like tail induced by the finite aperture of the mirrors (see also explanation in the end of Sec. II).

\subsubsection{Comparison with numerical clip approximation}

Numerical solution of the eigenvalue problem \eqref{eigenvalue} for a cavity with axially symmetric mirrors results in eigenfunctions (field distribution on the mirror) and diffraction loss. For the particular case of a cavity with spherical mirrors numerical results can be compared with both the analytical and numerical clip approximation. In what follows we compare i) the diffraction loss evaluated numerically solving the eigenvalue problem formulated in this paper \eqref{eigenvalue} and ii) the diffraction loss found from the clip approximation applied to a numerical solution of the eigenfunction problem for a lossless FP cavity.

The numerical clip approximation is formulated as follows. Using the function $\Psi_\ell$ found numerically at the spherical mirror surface of a lossless cavity we calculate the distribution after forward trip $\Psi_\ell^\text{forward}=\mathbf P^{(\ell)}_\text{forward}\Psi_\ell$ which is non-zero both at the mirror surface and outside of it. We then define the diffraction loss  as a fraction of energy flux ($\sim |\Psi_\ell^\text{forward}|^2$) propagating outside of the mirror and energy flux falling on the mirror (compare with \eqref{clip}):
\begin{equation}
      \label{clip3}
	\mathcal{L}_{clip} = 2\ \frac{\int\limits_{a_m}^{a}\big|\mathbf P^{(\ell)}_\text{forward}\Psi_\ell\big|^2xdx}{\int\limits_{0}^{a}\big|\mathbf P^{(\ell)}_\text{forward}\Psi_\ell\big|^2xdx}.
\end{equation}

We found that the difference between the diffraction loss calculated numerically from i) the clip approximation $\mathcal{L}_{clip}$ and, ii), the solution of the eigenvalue problem $\mathcal{L}$ is negligibly small. In particular, the parameter $\mathcal{L}_{clip}$ is smaller than the parameter $\mathcal{L}$ by no more than  a percent, as indicated by the results presented in Table~\ref{table1}. (This is not the case for clip approximation found from analytic consideration \eqref{clip2} with truncated Gaussian beam.)

Numerical values from Table~\ref{param} are utilized in the simulations. The axio-symmetric modes denoted as $AS0j$ are characterized with zero orbital momentum $\ell=0$, index $j$ stands for the radial mode number. The fundamental mode family has $j=0$. The dipole modes, denoted as $D1j$, are characterized with $\ell=1$ and radial mode number $j$. For axial symmetric mode

\begin{table}
		\caption{Diffraction loss (in ppm) for various modes of an aLIGO FP cavity (Table \ref{param}) found numerically using \eqref{eigenvalue} with number of points $N=512$ and window parameter \eqref{S} $S=2$ as well as and using the clip approximation \eqref{clip3}. The relative difference between the results of the calculations is less then $10^{-3}$.} \label{table1}
	\begin{tabular}{|c|c|c|c|c|c|c|c|}
		\hline
		  & AS00 & AS01 & AS02 & AS03 & D10 & D11 & D12\\
		\hline
		S & \multicolumn{4}{c}{$2.0029368342816$} & \multicolumn{3}{c}{$2.0009795519293$}\\
		\hline
		$\mathcal{L}$ & 0.40737 & 164.48 & 6202 & 99216 & 8.8913 & 1040.14 & 29688\\
		\hline	
		$\mathcal{L}_{clip}$ & 0.40699 & 164.40 & 6209 & 101810 & 8.8899 & 1040.18 & 29682\\
		\hline				
	\end{tabular}
\end{table}

It worth noting that the estimation of the round trip losses found with the analytical clip approximation \eqref{clip}  using the parameters listed in Table~\ref{param} gives $\mathcal L_0 \simeq 0.21$~ppm for the fundamental mode \eqref{clip2}. This is about two times smaller if compared with the numerically simulated losses presented in Table~\ref{table1} (0.46~ppm). The reason of this discrepancy will be studied elsewhere.

\subsubsection{Energy conservation test}

To verify the  validity of the numerical simulation we have to prove that the energy flux of light coming from one mirror ($\sim \int\big|\Psi_\ell\big|^2$) is equal to flux falling on both the opposite mirror and the area outside it (energy conservation law). It is convenient to introduce a divergence parameter $\sigma$ estimating our method accuracy
\begin{equation}
	\label{sigma_diverg}
	\sigma = \frac{\int\limits_{0}^{a}\big|\mathbf P^{(\ell)}_\text{forward}\Psi_\ell\big|^2xdx}{\int\limits_{0}^{a_m}\big|\Psi_\ell\big|^2xdx} - 1
\end{equation}

The parameter is zero in the ideal case. Our method has a limited simulation area and, hence, it is expected that (\ref{sigma_diverg}) slightly different from zero. We can neglect by this difference if it is much less than the diffraction loss. We found that in the case of a cavity with spherical mirrors with aLIGO parameters (Table \ref{param}) the divergence parameter $\sigma$ is smaller than the diffraction loss $\mathcal L$ by at least 3 orders for each mode within a FP cavity. Therefore, the proposed simulation technique is valid from the energy conservation law perspective.

\subsubsection{Accuracy dependence on the selected free parameters}

The accuracy of the numerical simulation depends on the on number of points $N$, selection of the coefficient ratio $\frac{P}{Q}$ as well as the window parameter $S$. We verify this dependence for an aLIGO FP interferometer (Table \ref{param}).
\begin{figure}
	\includegraphics[width=0.49\textwidth]{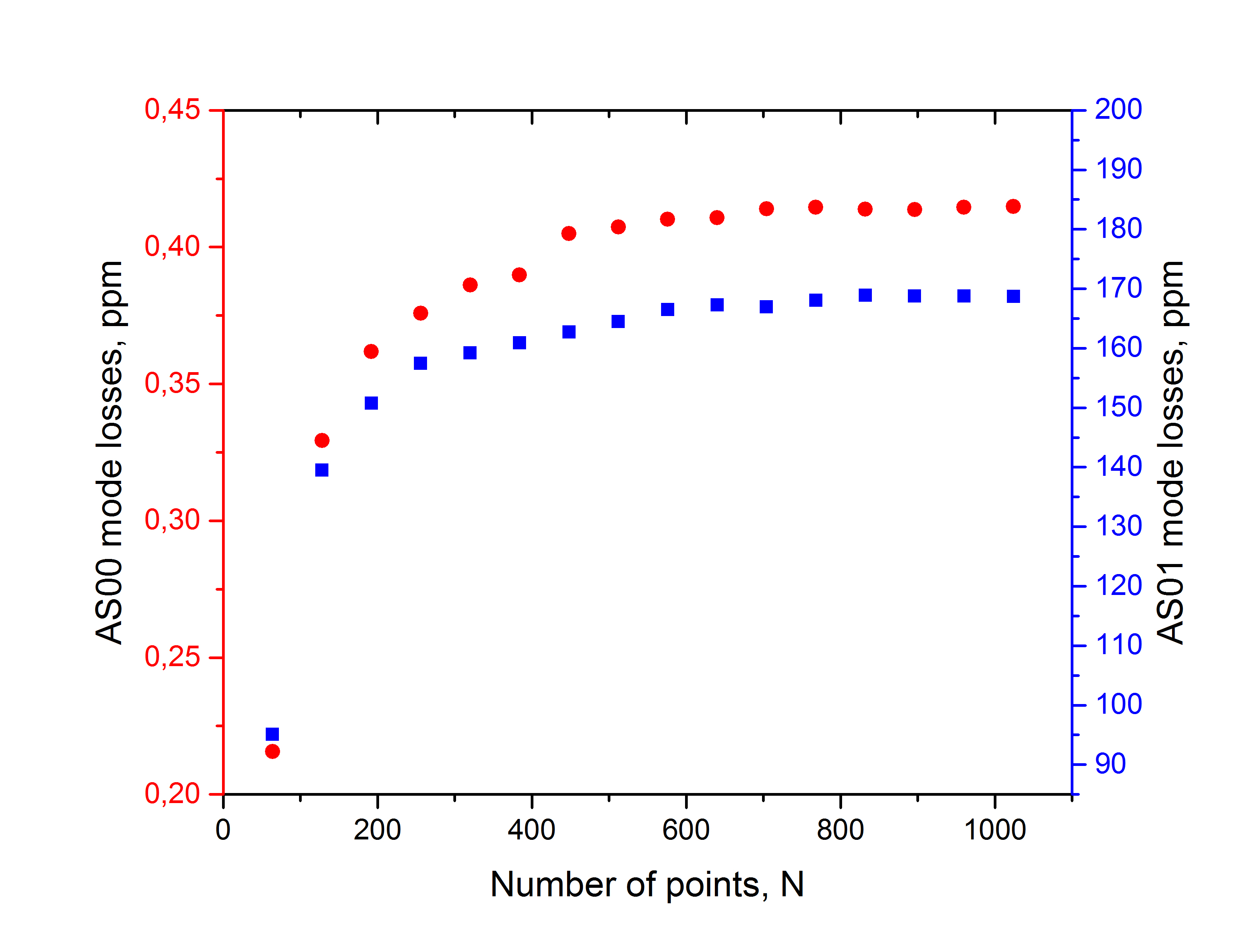}
	\caption{Dependence of the numerically found diffraction loss of the axial symmetric modes AS00 and AS01 on the quantization number $N$. The physical parameters were taken from Table \ref{param}. The free parameters \popl{$S = \frac{\xi_{N}}{\xi_{N/2}} \simeq 2$} and $P/Q = 1$ were selected.}\label{L(N)}
\end{figure}
As shown in Fig.~(\ref{L(N)}) the observed difference between the attenuation values found in cases of $N = 512$ and $N = 1024$ is less than $4\%$.
The simulation time increases as $O(N^{3})$ \cite{wiki}, but the change of the number $N$ by 10\% results in variations of the simulated attenuation by less than a 1\%. Hence the number $N = 512$ is good enough for the majority of calculations for the cavity with selected parameters.

\subsubsection{Window parameter $S$ and ghosts solutions}\label{window}

\begin{figure}[h]
 	\includegraphics[width=0.49\textwidth]{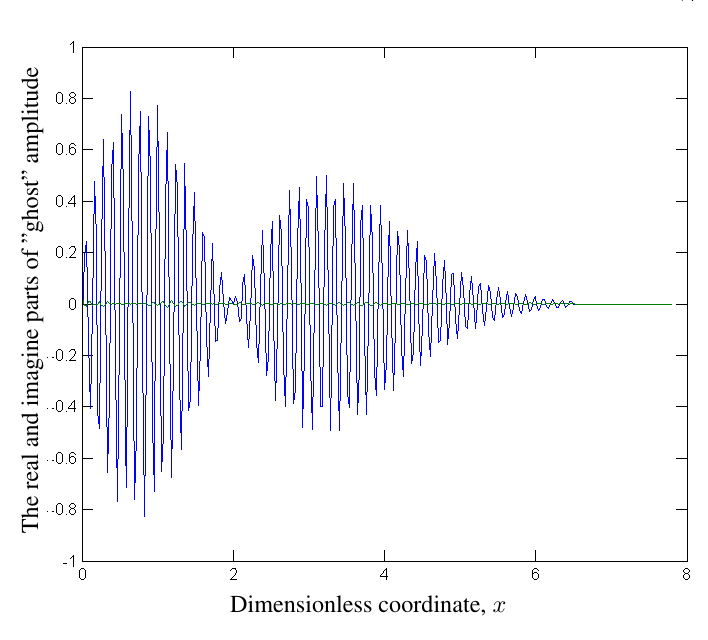}
	\caption{Unphysical solutions of the eigenvalue problem (''ghosts''). Blue dots stand for the real part of the solution, green dots imagine part of them (close to horizontal axis). As we can see envelopes of such solutions are similar to profile of axiosymmetric and dipole modes. Losses of them are on the order of $10^{-2}$ ppm.}\label{ghost}
\end{figure}
The numerical solution of the eigenvalue problem (\ref{eigenvalue}) results in a number of solutions with non-physically small attenuation. These ghost solutions are characterized with unrealistic oscillations of the mode amplitude along the radial coordinate. The envelope of the oscillations is identical to the profile of the corresponding Laguerre-Gaussian mode of the cavity (see example in Fig.~\ref{ghost}).

\begin{figure}[h]
 	\includegraphics[width=0.22\textwidth]{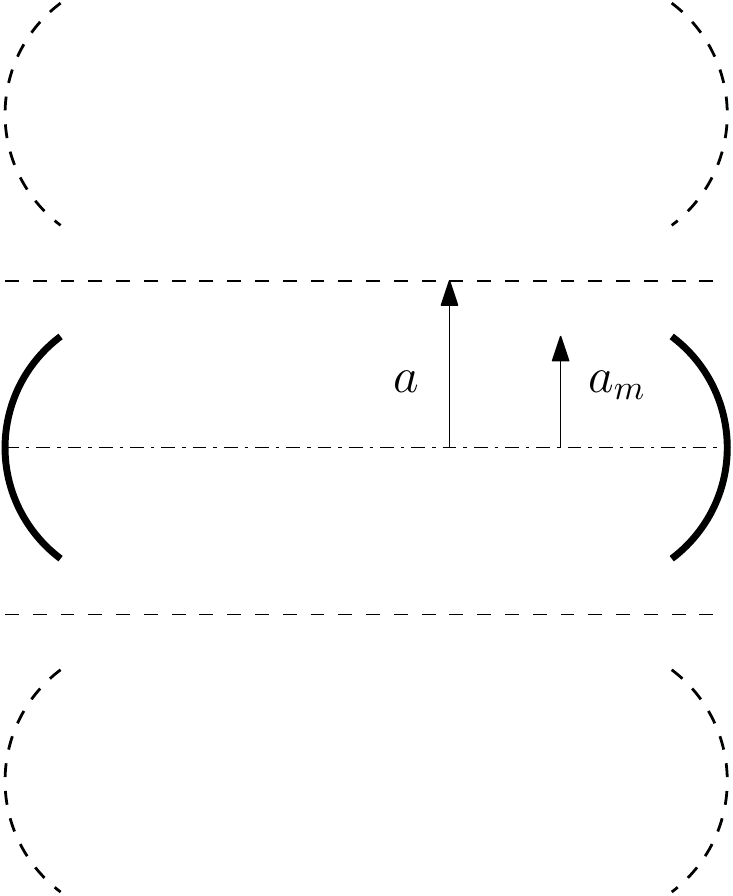}
	\caption{Calculation of FP cavity using Fourier transform means that there are alias cavities around shown by dashed lines.}\label{alias}
\end{figure}

For the window parameter $S\simeq 2$ \eqref{S} we have seen about 30 ghosts with relatively small loss. The number of these ghosts dramatically decreases for a larger window parameter ($S>5$). The  most probable reason of the ghost solutions is the numerical Fourier transform leading to appearance of the high frequency components in the spectrum \cite{Nyquist1928, kotelnikov1933e, shannon1948}, caused by unfulfillment of Nyquist--Shannon--Kotelnikov sampling theorem conditions.
Recall Hankel is based on Fourier transform \eqref{FresnelFT} and it has the same aliasing problem as the FFT-based numerical calculation. When one FP cavity is calculated, there are alias FP cavities around as shown on Fig.~\ref{alias}. For values $S\gg 1$ the influence of alias is negligible, where as for $S\simeq 1$ influence of alias cavities is strong, because large part of light from alias cavities return. 

On the other hand, usage of a large window parameter $S$ has an obvious disadvantage because of the corresponding small number of points located at the mirror surface and, hence, bad accuracy of the simulation. Really, an increase of the window parameter $S$ at a fixed total number of points $N$ leads to the reduction of the effective number $N_{mir}$ of points falling on the mirror in accordance with
\begin{equation}
\label{Nmir}
N_{mir} \simeq \frac{a_m }{a}\cdot N = \frac N S.
\end{equation}
It is more convenient to cope with the oscillating solutions by setting a low-pass spatial filter removing the ghosts.
\begin{figure}[h]
\includegraphics[width=0.45\textwidth]{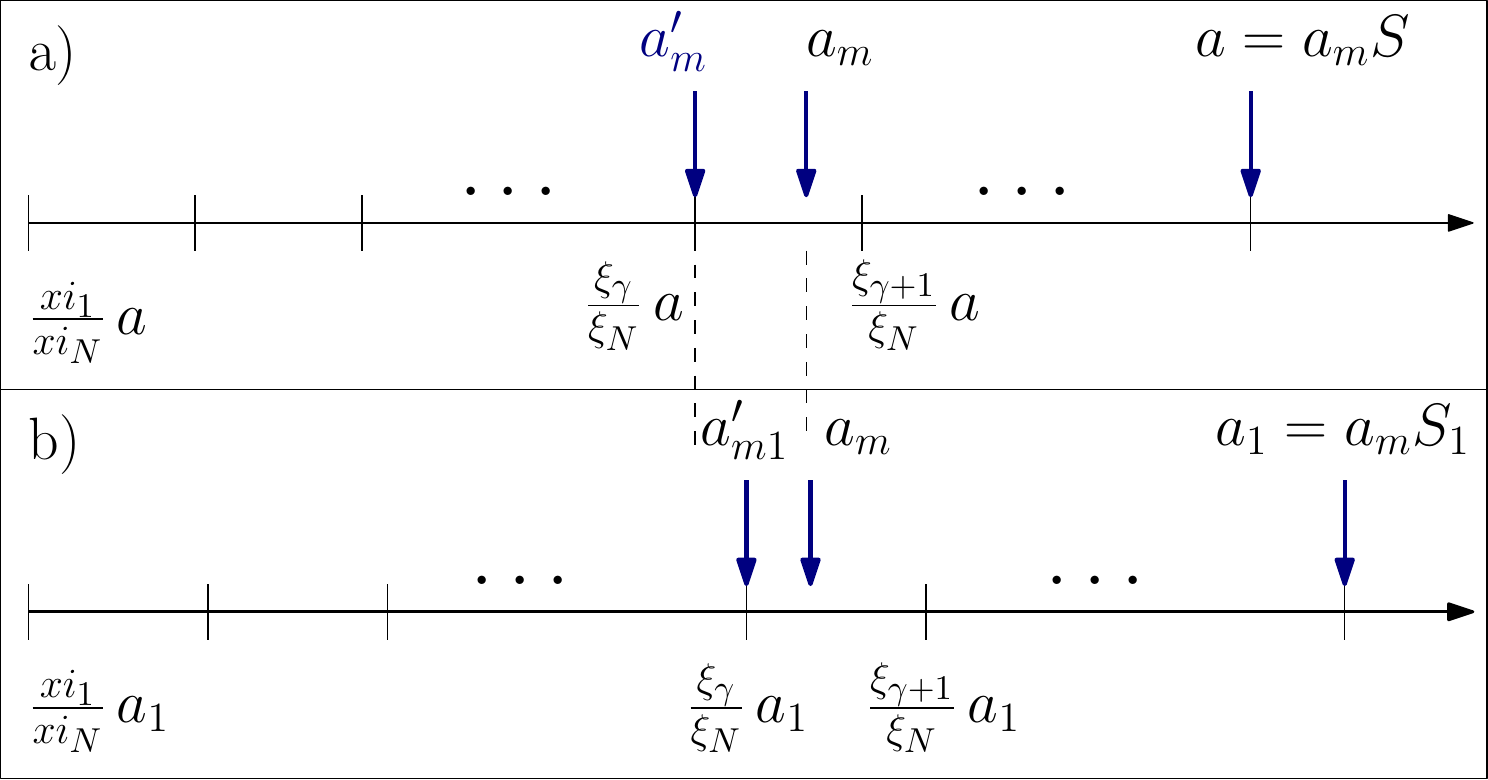}
\caption{a) Window parameter S should be define as (\ref{S_ratio}) to avoid error in calculations due to mismatch between effective radius $a_m'$ of mirror (defining diaphragm function $D_k$ (\ref{diaphragm})) but not radius $a_m$ of mirror (\ref{S}). b)  The same radius $a_m$ and slightly larger window parameter $S_1>S$.}\label{wrongS}
\end{figure}
\begin{figure}[h]
 	\includegraphics[width=0.49\textwidth]{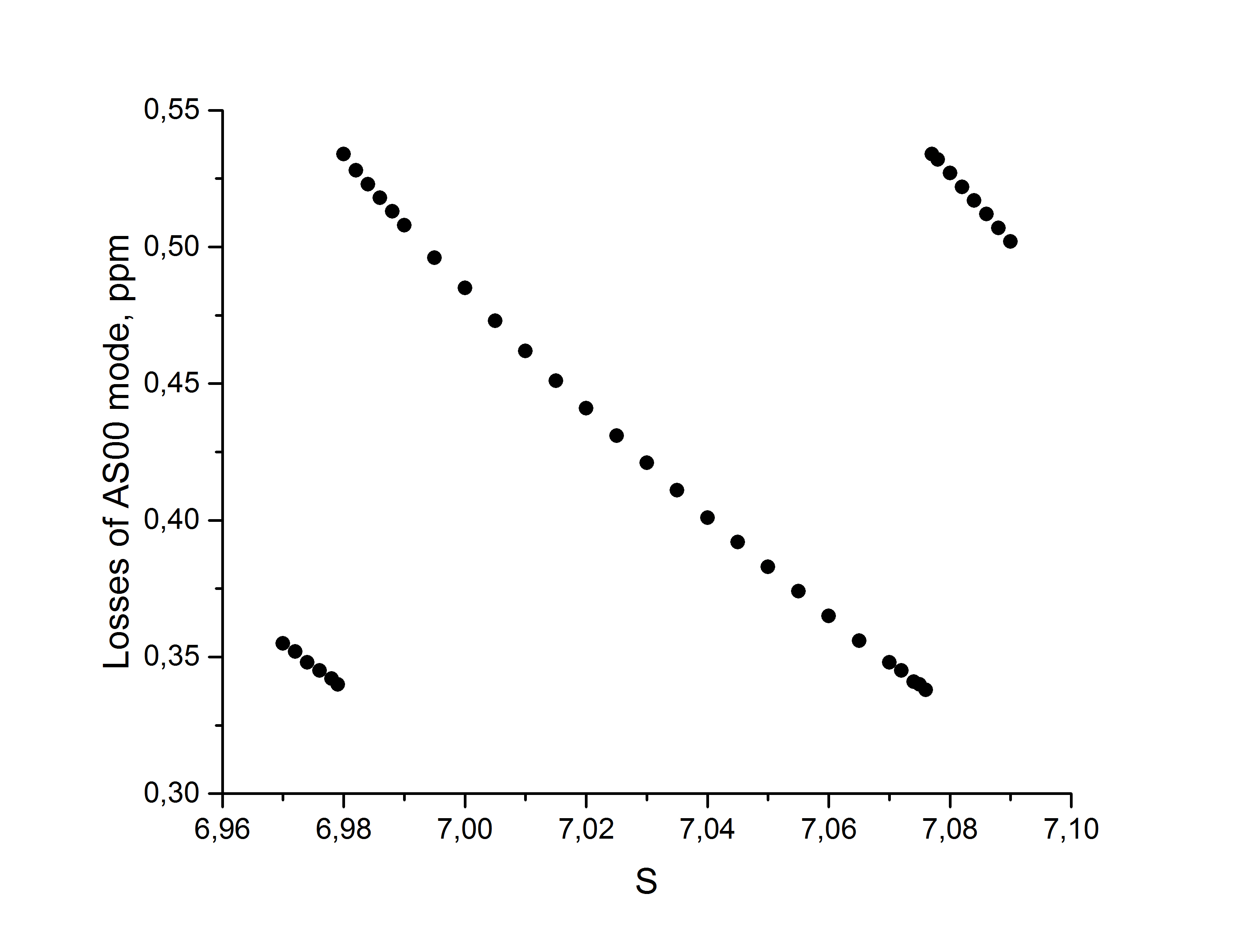}
	\caption{Influence of small variations of the window parameter S on simulated numerically diffraction loss of the AS00 mode. Here the aLIGO parameters (Table \ref{param}) along with $N = 512$ and $P/Q = 1$ are utilized.}\label{L(big_S)}
\end{figure}

An arbitrary selection of the value of the parameter $S$ may lead to an unwanted effect of branching of the attenuation value found numerically. The results of a numerical simulation of the diffraction loss as a function of window parameter $S$ are shown in Fig.~\ref{L(big_S)}. We see that oscillations of the loss value reaches almost 30$\%$. Let us stipulate how to use the window parameter $S$ correctly to avoid the oscillations.

We select the mirror radius $a_m$ with parameter $S$ and find the diaphragm radius $a = a_m S$. The parameter $a $ coincides with root $\xi_N$ whereas $a_m$ may not coincide with any root $\xi_j$. For instance, let $a_m $ be localized between two discrete points $\frac{\xi_\gamma}{\xi_N}a < a_m < \frac{\xi_{\gamma+1}}{\xi_N}a$). It means, that effective mirror's radius $a_m' = \xi_\gamma a/\xi_N$ is {\em smaller} than $a_m$ (see Fig.~\ref{wrongS}a). Hence, the {\em effective} window parameter $S'$ is {\em larger} than $S$
 \begin{equation}
	\label{Sratio}
	S' = \frac{\xi_N}{\xi_{\gamma}} >S
\end{equation}
If we increase window parameter from $S$ to $S_1$ a tiny bit, the both the diaphragm $a_1$ and effective radius of the mirror $a_{m1}'>a_m'$ also increase as shown on Fig.~\ref{wrongS}b.

Obviously, the diffraction loss value found by the simulation depends not on the radius $a_m$ of mirror but on the effective radius $a_m'$ (or $a_{m1}'$). The larger is the effective radius the smaller is the diffraction loss. Hence, the diffraction loss should be {\em smaller} for the {\em larger} window parameter ($S_1$) than for the smaller one ($S$).   Figure~\ref{L(big_S)} illustrates this statement.

To avoid this problem one has to always select the window parameter correctly so that the mirror radius $a_m$ coincides with a discrete point belonging to the set $\xi_i$  in accordance with the rule
\begin{equation}
	\label{S_ratio}
	S = \frac{\xi_N}{\xi_{\gamma}}.
\end{equation}
Such a selection corresponds to the {\em bottom} edge of the curve of Fig.~\ref{L(big_S)}.

\begin{figure}
	\includegraphics[width=0.49\textwidth]{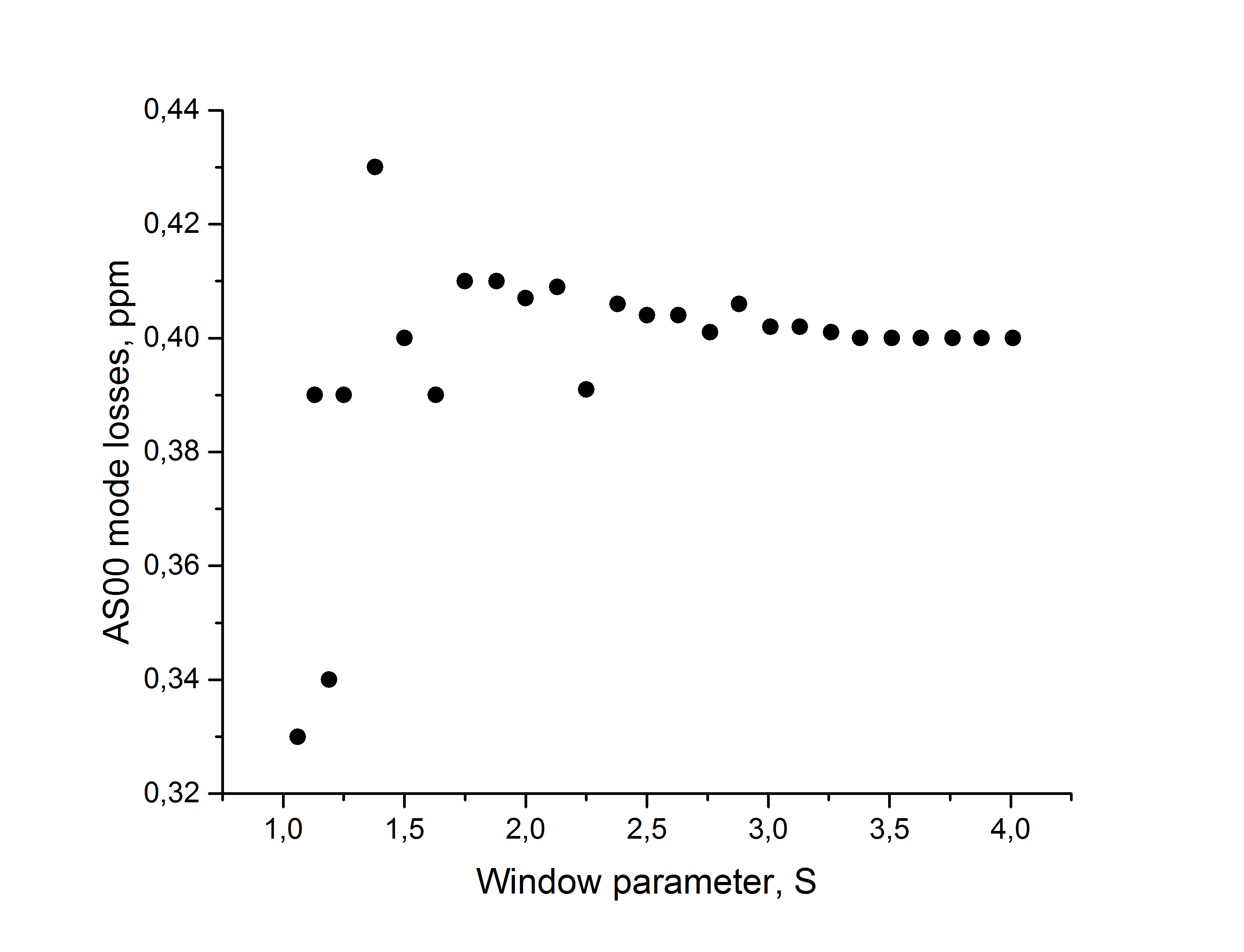}
	\caption{Influence of the window parameter $S$ on the numerically simulated value of the diffraction loss of the mode AS00 for the fixed number $N_{mir}=256$ of points on mirror. Here we used aLIGO parameters of cavity (Table \ref{param}) with $P/Q = 1$. Mode loss vanishes at $S = 1$ because points describing tail of the mode are not taken into account.}\label{L(S)1}
\end{figure}

\begin{figure}
	\includegraphics[width=0.49\textwidth]{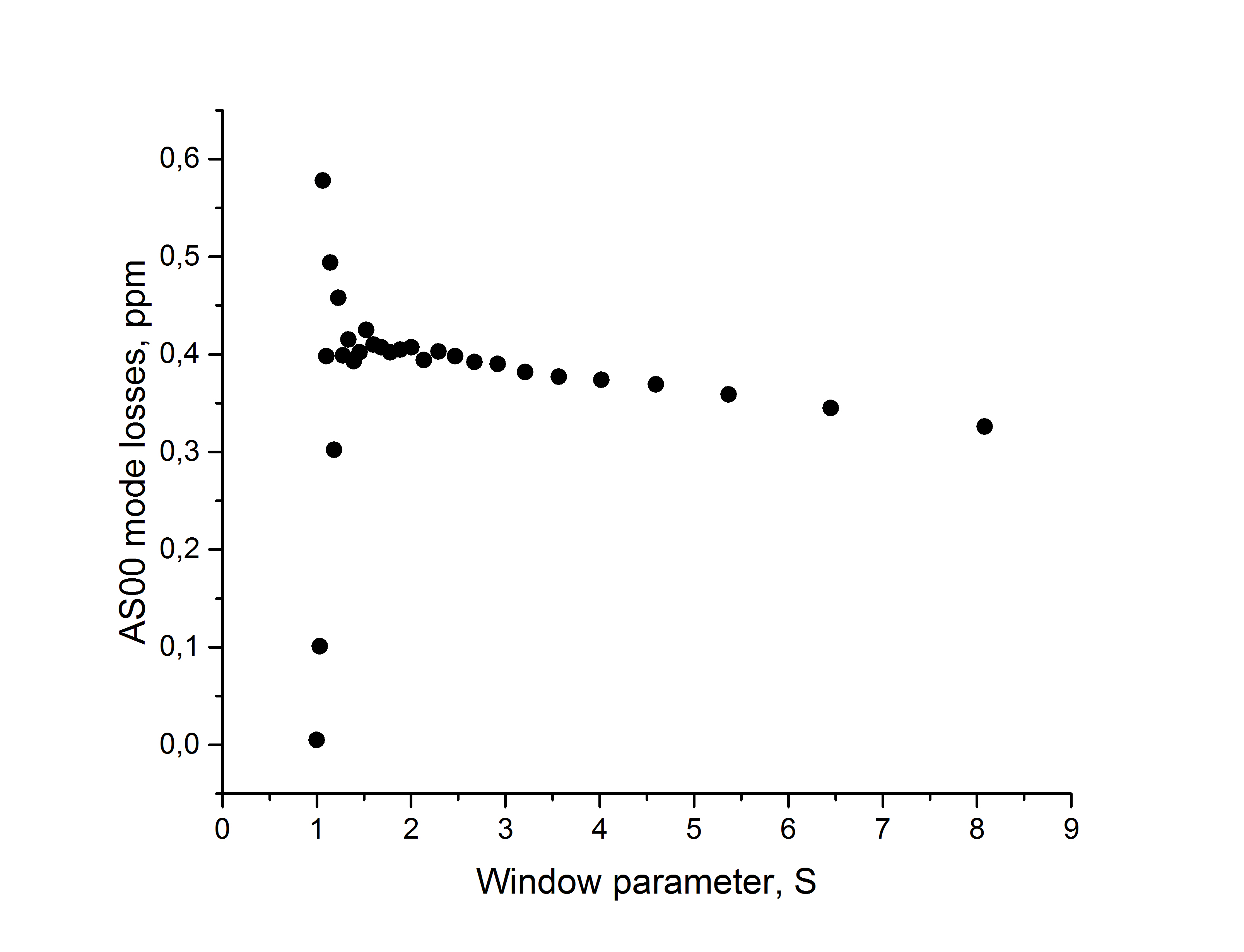}
	\caption{A dependence of the simulated numerically diffraction loss of the mode AS00 on the parameter $S$. The loss is found for the fixed number $N = 512$ of the points in Fourier space. Here we used aLIGO parameters of cavity (Table \ref{param}) with $P/Q = 1$. }\label{L(S)2}
\end{figure}

There exists a finite set of possible values of $S$ is we adopt Eq.~(\ref{S_ratio}). A dependence of the simulated numerically diffraction loss of the mode AS00 on parameter $S$ for the fixed $N_{mir} = 256$ (\ref{Nmir}) and properly selected $S$ (see formula (\ref{S_ratio}) and Fig. \ref{L(S)1}) as well as for the fixed number of points in Fourier space $N = 512$ (Fig. \ref{L(S)2}) shows the general accuracy limitation of the simulation technique as the result depends on the $S$ selection.
The variation of the $S$ value leads to fluctuations of the mode loss reaching not more 5$\%$ near the point $S \simeq 2.5$ (instead of 30$\%$ when $S$ is selected inappropriately, as shown in Fig.\ref{L(big_S)}). The parameter $2.5$ corresponds to the optimal distribution of the number of points corresponding to the body and the tail of the mode. The simulated loss increases for the large $S$ since the number of points covering the body of the mode decreases. Extensive tests show that $S = 2\div 3$ and $N \ge 512$  are the most suitable values for simulations of the aLIGO FP cavity with acceptable calculation time. Simulations of cavities of different structure requires optimization. The choice of the ratio $P/Q$ does not impact the simulation accuracy. For instance, the diffraction loss for the parameter ranging from $2$ to $1/2$ differs no more than 1$\%$ for different HOOM.

\section{Sensitivity to Small Tilts}\label{SmallTilt}

Practical applications of FP cavities with non-spherical mirrors \cite{92pare,05kuznetsov,07tiffani} often call for an evaluation of the stability of the cavity with respect to small tilt of the mirrors \cite{04DaShStVyTh, 04ShStVy, 14a1FeDeVyMa, 16a1MaPoYaVy}. A direct simulation of the cavity with tilted mirror is hindered by the asymmetrical morphology of the system.

Let us assume that the end mirror is tilted by a small angle $\theta$, as shown in Fig.~\ref{geomYa}. The eigenvalue problem is formulated similarly to Eq.~(\ref{Fresnel2}) by taking into account the phase shift $\beta$ due to the tilt
\begin{align}
\label{tildePsi}
	\widetilde{\lambda}^2_{n, \ell}\widetilde{\Psi}_{n, \ell} (\vec{x}_1) = \int g(\vec{x}_1, \vec{x}_2)e^{2i\beta}\widetilde{\Psi}_{n, \ell} (\vec{x}_2)d\vec{x}_2,
\end{align}
where $g(\vec{x}_1, \vec{x}_2)$ is a propagator of non-perturbed problem and phase shift  $\beta \equiv \theta k r \cos (\phi) = \theta \sqrt {kL}\, x_2 \cos (\phi)$ is produced by the mirror tilt, $\phi$ is azimuthal angle.

Since the tilt angle is small we use expansion $e^{2i\beta} \simeq 1 + 2i\beta - 2\beta^2+... \equiv 1 + \delta(\vec{x}_2)$. Also we assume that the parameters of a mode of such a perturbed cavity $\widetilde{\Psi}_{n, \ell} (\vec{x})$ can be expressed as an expansion over modes of the non-perturbed cavity $\Psi_{n, \ell}(\vec{x})e^{i\ell\phi}$ in a form
\begin{equation*}
	\widetilde{\Psi}_{n, \ell} (\vec{x}) = 	C_{n\ell}\Psi_{n, \ell}(\vec{x})e^{i\ell\phi} + \sum_{k \neq n, m \neq \ell} C_{km}\Psi_{n, m}(\vec{x})e^{im\phi}
\end{equation*}
The method of successive approximations allows evaluating the value $\tilde {\mathcal L}_{00}$ of the fundamental axial symmetric mode for cavity
\begin{align}
\label{tildeL}
 \tilde {\mathcal L}_{00} &\simeq   \mathcal{L}_{00}\left[1 + \left(\frac{\theta}{\theta_\text{perm}}\right)^2\right],
      \quad \frac{1}{\theta^2_\text{perm}}= \frac {kL S_U}{\mathcal L_{00}} \\
 \label{SU}
 S_U &\equiv \Re\left[U_{00,00} - 2\sum_{k \in \mathbf{Z}_+}\frac{ \lambda_{k1}^2 \big|U_{k1,00}\big|^2}{\lambda_{k1}^2 -\lambda_{00}^2 }\right],\\
	U_{00,00} &\equiv   \int \psi^*_{00}(x_2)\psi_{00 }(x_2)\, x_2^3\, dx_2,\\
	U_{k1,00} &\equiv   \int \psi^*_{k1 }(x_2)\psi_{00 }(x_2)\, x_2^2\, dx_2,
\end{align}
here $\mathbf{Z}_+$ is a set of integer non-negative numbers. See details in Appendix \ref{app1}. The sense of $\theta_\text{perm}$ is simple: at $\theta= \theta_\text{perm}$ round trip loss increases by 2 times.

For parameters listed in Table~\ref{param} we have found an estimate for tilt angle $\theta_{perm}$ \eqref{tildeL} which doubles diffraction loss of the fundamental optical mode AS00 of the FP cavity
\begin{align}
 \theta_\text{perm} \simeq 0.6\cdot 10^{-6}.
\end{align}
Note that the estimation \eqref{thetaGeom}, obtained from the geometrical analysis, gives angle $\theta^\text{geomG}_\text{perm}$ about two times smaller. Whereas estimation \eqref{thetaSAG}, obtained from successive approximation using truncated Gaussian modes, gives $\theta^\text{saG}_\text{perm}$ about 2~times larger. The estimate \eqref{tildeL} based on the numerically calculated mode profiles seems more reliable  as compared with estimates based on truncated Gaussian distributions (formally valid  for infinite mirrors only).

\section{Sensitivity to small roughness}

\begin{figure}
	\includegraphics[width=0.49\textwidth]{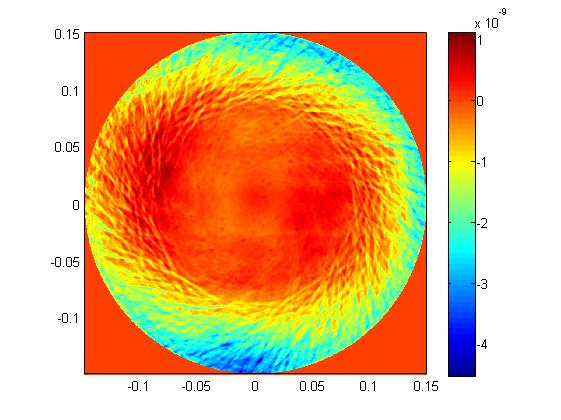}
	\caption{A typical roughness of spherical profile of an aLIGO mirror measured in meters after tilt and curvature subtraction using a standard mathematical approach \cite{roughness}.}
	\label{rough1}
\end{figure}

Roughness of the mirror surface is another reason of increase of the loss of the optical modes of a realistic FP cavity. The problem can be handled in a way similar to the case of the tilted mirror (\ref{tildePsi}), but the phase shift radial dependence $\beta(r, \varphi)$ in this case is arbitrary (Fig.\ref{rough1}).
A method of successive approximations allows to find the influence of the roughness on the eigenvalue of the fundamental optical FP mode (AS00)  and evaluate the increase of the loss of the mode due to the nonideality of the cavity mirror. We assume that the roughness is small is compared with $\lambda/F$, where $F$ is the finesse of the cavity and write
\begin{align}
	e^{-2i\beta} &= \delta_1 + \delta_2 + \delta_3 +\dots, \\	
	\tilde\lambda_{00}^2 &= \lambda_{00 }^2 + \left(\tilde\lambda_{00 }^2\right)^{(1)} + \left(\tilde\lambda_{00}^2\right)^{(2)}+\dots,\\
	\left(\tilde\lambda_{00 }^2\right)^{(1)}  &= \lambda_{00}^2 V^{(1)}_{00,00 }, \\
	V_{00,00}^{(1)}&\equiv \int \psi^*_{00 }(x_1) \,\delta_1(\vec x_1)\, \psi_{00 }(x_1) \,x_1\,d x_1\, d\phi_1 ,\\
	\left(\tilde\lambda_{00 }^2\right)^{(2)}  &= \lambda_{00}^2\Delta V,\\
	 \Delta V \equiv & \left( V^{(2)}_{00,00 } -  2\sum_{
	 k, m \in \mathbf{Z}_+}  \frac{\lambda_{km}^2}{\lambda_{00}^2 - \lambda_{km}^2}\cdot\left| V_{km,00}^{(1)} \right|^2\right),
	 \label{deltaV}
\end{align}
where $\lambda_{km}^2$ are the unperturbed eigenvalues of the cavity.

It is convenient to tune the system so that $V_{00, 00}^{(1)}=0$ by selecting the zero averaged level of roughness and then evaluate all the other matrix elements relatively this level. Details of the calculations are discussed in Appendix \ref{app2}. The perturbed by toughness roundtrip loss may be expressed as $\tilde {\mathcal L}_{00} = 1 - \big|\tilde\lambda_{00}^2\big|^2$ using Eq.~(\ref{deltaV}).

We evaluated the perturbed value of loss for the cavity mirrors characterized with various roughness maps $\beta(r, \varphi)$ of aLIGO mirrors \cite{roughness} and found that the roughness does not increase the loss by more than 3~ppm (for the roughness map shown on Fig.~\ref{rough1} the added loss is about 1.6 ppm). The added loss value does not depend on variation of the mirror shape \cite{16a1MaPoYaVy} due to similarity of the profile of the fundamental modes of the FP cavities. We are performing a more detailed study with the goal to figure out the source of the excessive loss in a realistic aLIGO cavity and the results will be published elsewhere. 

\section{Conclusion}

We have performed a detailed study of diffraction loss of a realistic Fabry Perot cavity. A method for numerical analysis of both the eigenmodes and {\em complex} eigenvalues of high finesse Fabry-Perot cavities assembled by axial symmetric mirrors with arbitrary profile and {\em finite} size is proposed and described in detail. This method is efficient for both the fundamental and high order optical modes of the cavity. We use the method to find the finesse of the cavity utilizing parameters of aLIGO interferometer for numerical estimations.

Only radial dependence of the field distributions on mirror needs to be evaluated in our approach. It takes much smaller time if compared with a direct solution of a 2D problem. We show that our method can be utilized to find loss of a FP cavity with small non-axial asymmetry perturbation of its mirrors, in particular, for evaluation of diffraction loss of a Fabry-Perot cavity with tilted mirror and mirrors with small roughness. The technique could be useful for explanation of the observed in experiment round trip loss in the aLIGO interferometers as it allows estimation of the attenuation due to excessive scattering as well as mode mismatch in the cavities. This study will be published elsewhere.

\acknowledgments

LIGO was constructed by the California Institute of Technology and Massachusetts Institute of Technology with funding from the National Science Foundation, and operates under cooperative agreement PHY-0757058.
M.P. and S.V. acknowledge support from the Russian Foundation for Basic Research (partially, Grant No. 16-52-10069), Russian Science Foundation (researches in Section IV supported by Grant No. 17-12-01095), Russian Science Foundation (partially, grant No.~17-12-01095) and National Science Foundation (researches in Sec. V, VI supported by Grant No. PHY-130586).

\appendix

\section{Derivation of Eq. \eqref{Fresnel2}}\label{Bessel}

Here we derive equation \eqref{Fresnel2} from \eqref{Fresnel1}. Let write \eqref{Fresnel1} in polar coordinates using presentations \eqref{Fresnel1}:
\begin{subequations}
 \label{Fresnel3}
   \begin{align}
   \Psi_1(x_1)&\,e^{i\ell\phi_1} = -\frac{i}{2\pi}\int_0^\infty  x_2\, d x_2 \int_0^{2\pi}d\phi_2 \times\\
   \times & \exp\left(i\left[\frac{|x_1^2 + x_2^2 - 2x_1x_2\cos(\phi_1-\phi_2)}{2}\right]\right)\times\nonumber\\
    &\quad \times \Psi_2(x_2)\, e^{i\ell\phi_2}\,
   \end{align}
\end{subequations}
Then we multiply both sides of \eqref{Fresnel3} by $e^{-i\phi_1}$ and integrate over variable $\eta =\phi_2-\phi_1$ using known integral formula for Bessel function (for example, see formula 21.8.18 in \cite{68Korn}):
\begin{align}
   \label{BesselInt}
 \int_{-\phi_1}^{2\pi-\phi_1} e^{i\big(\ell \eta  +x_1x_2\cos \eta\big)}\, d\eta = 2\pi\, i^\ell \, J_\ell(x_1x_2)\,.
\end{align}
The integral \eqref{BesselInt} does not depend on $\phi_2$ because under integral function has period $2\pi$.

After substitution we obtain \eqref{Fresnel2}.

\section{Successive Approximation Method for Calculations of Tilt Stability}\label{app1}

After multiplying \eqref{tildePsi} by $\Psi_{00}^*(\vec{x}_1)$ or by $\Psi_{nl}^*(\vec{x}_1)e^{-im\varphi_1}$ and averaging over $d\vec{x}_1$ we obtain set of equations:

\begin{align}	
	\left(\widetilde{\lambda}_{00}^2 - \lambda_{00}^2\right) & C_{00} = \lambda_{00}\sum_{k, m \in \mathbf{Z}_+}\lambda_{km}C_{km}V_{00,km} \\
	\left(\widetilde{\lambda}_{00}^2 - \lambda_{nl}^2\right) & C_{nl} = \lambda_{nl}\sum_{k, m  \in \mathbf{Z}_+}\lambda_{km}C_{km}V_{nl,km},\\
	V_{nl, km} \equiv \int & \Psi^*_{nl}(\vec{x}_1)e^{-il\varphi_2}\delta(\vec{x}_1)\Psi_{km}(\vec{x}_1)e^{im\varphi_1}d\vec{x}_1
\end{align}

For main axial symmetric AS00 mode we apply  method of successive approximations for calculation matrix elements $V_{00, km}$. Expanding in series $\delta(\vec{x}) \simeq \delta^{(1)} + \delta^{(2)} + \dots$ we carried out:

\begin{align*}
	\delta^{(1)} &= 2i\beta  =  i\theta\sqrt{kL}\, x_2 \left( e^{i\phi_2} +e^{-i\phi_2}\right),\\
    \delta^{(2)} &=   -\frac{1}{2}\theta^2kLx_2^2\left(2+e^{i\ell 2\phi_2}+ e^{-i\ell 2\phi_2}\right),\\
	V_{00,km}^{(1)} &= \delta_{(\pm 1) m}\, 2\pi i\,\theta \sqrt{kL}\,\int \psi^*_{00 }(x_2)\, \psi_{k1 }(x_2) \,  x_2^2\, d x_2,\\
	V_{00,km}^{(2)} &= - \theta^2 kL \Big(\delta_{m0}\, 2\pi \int \psi^*_{00 }(x_2)   \psi_{k0 }(x_2)\, x_2^3\, d x_2 \\
		 &+ \delta_{m(\pm 2)}\, 2\pi\, \frac 1 2 \int \psi^*_{00 }(x_2)   \psi_{k2}(x_2)\, x_2^3\, d x_2 \Big).
\end{align*}

The next step is expansion in series over successive orders of smallness:
\begin{align*}
\tilde \lambda_{00}^2 &= \left(\tilde \lambda_{00}^2\right)^{(0)} +  \left(\tilde \lambda_{00}^2\right)^{(1)} +
    \left(\tilde \lambda_{00}^2\right)^{(2 )}+\dots,\\
 C_{mn} &=C_{mn}^{(0)} + C_{mn}^{(1)} +  C_{mn}^{(2)}+\dots
\end{align*}

We put at initial approximation of zero order of the smallest parameter $\theta$: $C_{00}^{(0)} = 1 = \delta_{0m}\delta _{0n},\quad C_{n\ell}^{(0)} =0$. So it is easy to get eigenvalue of our problem:

\begin{align*}
 \sim \theta^0 :&\quad \left(\tilde \lambda_{00}^2\right)^{(0)}
    =\lambda_{00}^2,\\
    \sim \theta^1 :&\quad \left(\tilde \lambda_{00}^2\right)^{(1)} = \lambda_{00}^2V_{00,00}^{(1)}\equiv 0,\\
    \sim \theta^2 :&\quad \left(\tilde \lambda_{00}^2\right)^{(2)}
	  = \lambda_{00}^2 \left(V_{00,00}^{(2)} - 2\sum_{k \in \mathbf{Z}_+}\frac{ \lambda_{k1}^2 \big|V_{k1,00}^{(1)}\big|^2 }{\lambda_{00}^2 - \lambda_{k1}^2}\right)
\end{align*}

Summing up the first three approximations of eigenvalue:
\begin{align*}
  \tilde \lambda_{00}^2  =  \lambda_{00}^2\left(1 +V_{00,00}^{(2)} - 2\sum_{k \in \mathbf{Z}_+}\frac{ \lambda_{k1}^2 \big|V_{k1,00}\big|^2}{\lambda_{00}^2 - \lambda_{k1}^2}\right)
\end{align*}
where

\begin{align*}
	\big|V_{00,km}^{(1)}\big|^2 \equiv - \delta_{(\pm 1) m}\theta^2 kL \left|2\pi\int \psi^*_{k1 }(x_2)\psi_{00 }(x_2)\, x_2^2\, dx_2\right|^2
\end{align*}

In terms of losses which are equal to ${\mathcal L}_{00} = 1 - \big|\lambda_{00}\big|^2$ we can rewrite this expression as

\begin{align}
  \tilde {\mathcal L}_{00} =  \mathcal L_{00} + \Re\left[-V_{00,00}^{(2)}+2\sum_{k \in \mathbf{Z}_+}\frac{ \lambda_{k1}^2 \big|V_{k1,00}\big|^2}{\lambda_{00}^2 - \lambda_{k1}^2}\right] +\dots
\end{align}
$\mathbf{Z}_+$ is set of non-negative integer numbers.

From this formula we obtain \eqref{tildeL} and \eqref{SU}. Note, the multiplier 2 in \eqref{SU} appears due to two terms in sum. It corresponds to account of dipole modes $e^{\pm i\varphi}$.

\section{Successive Approximation Method for Calculations of Roughness Stability}\label{app2}

Using analogical approach as in Appendix \ref{app1} and taking into account that roughness of specular surface $\beta(r, \varphi)$ has an angular dependence we obtained that

\begin{align}	
	\left(\widetilde{\lambda}_{00}^2 - \lambda_{00}^2\right) & C_{00} = \lambda_{00}\sum_{k, m \in \mathbf{Z}_+}\lambda_{km}C_{km}V_{00,km} \\
	\left(\widetilde{\lambda}_{00}^2 - \lambda_{nl}^2\right) & C_{nl} = \lambda_{nl}\sum_{k, m \in \mathbf{Z}_+}\lambda_{km}C_{km}V_{nl,km},\\
	V_{nl, km} \equiv \int & \Psi^*_{nl}(\vec{x}_1)e^{-il\varphi_2}\delta(\vec{x}_1)\Psi_{km}(\vec{x}_1)e^{im\varphi_1}d\vec{x}_1
\end{align}

Expanding in series $e^{-2i\beta} \equiv \delta(x, \varphi) \simeq \delta^{(1)} + \delta^{(2)} + \dots$ and assuming smallness of $\beta$ we carried out that:
\begin{align}
	V_{00,ml}^{(1)}&\equiv \int \Psi^*_{00}(x_1) \,\delta_1(\vec x_1) \Psi_{ml}(x_1) e^{-i\ell\varphi}\,x_1\,d x_1\, d\phi_1 \\
	V_{00,00}^{(2)}&\equiv \int \Psi^*_{00}(x_1) \,\delta_2(\vec x_1) \Psi_{00}(x_1)\,x_1\,d x_1\, d\phi_1
\end{align}

Further we expanded eigenvalue in series over successive orders of smallness:
\begin{align*}
\tilde \lambda_{00}^2 &= \left(\tilde \lambda_{00}^2\right)^{(0)} +  \left(\tilde \lambda_{00}^2\right)^{(1)} +
    \left(\tilde \lambda_{00}^2\right)^{(2 )}+\dots,\\
 C_{mn} &=C_{mn}^{(0)} + C_{mn}^{(1)} +  C_{mn}^{(2)}+\dots
\end{align*}

We put at initial approximation of zero order of the smallest parameter $\theta$: $C_{00}^{(0)} = 1 = \delta_{0m}\delta _{0n},\quad C_{n\ell}^{(0)} =0$. And then:

\begin{align*}
 \sim \beta^0 :&\quad \left(\tilde \lambda_{00}^2\right)^{(0)}
    =\lambda_{00}^2,\\
    \sim \beta^1 :&\quad \left(\tilde \lambda_{00}^2\right)^{(1)} = \lambda_{00}^2V_{00,00}^{(1)},\\
    \sim \beta^2 :&\quad \left(\tilde \lambda_{00}^2\right)^{(2)}
	  = \lambda_{00}^2 \left(V_{00,00}^{(2)} - 2\sum_{k,m  \in \mathbf{Z}_+}\frac{ \lambda_{km}^2 \big|V_{km,00}^{(1)}\big|^2 }{\lambda_{00}^2 - \lambda_{km}^2}\right)
\end{align*}

Summing up three steps of approximation of eigenvalue, we get:
\begin{align}
  \label{deltaL}
  \tilde \lambda_{00}^2  =  \lambda_{00}^2\left(1 + V_{00, 00}^{(1)} + V_{00,00}^{(2)} - 2\sum_{k, m \in \mathbf{Z}_+}\frac{ \lambda_{km} \big|V_{km,00}\big|^2}{\lambda_{00} - \lambda_{km}}\right)
\end{align}

We want to note that presence of the multiplier 2 in relations (\ref{deltaV}, \ref{deltaL}) is due to accounting two conjugate modes with angular dependence $e^{\pm im \varphi}$.

\end{document}